\documentclass[12pt,onecolumn]{IEEEtran}
\usepackage{amsmath,amsfonts}
\usepackage{array}
\usepackage[caption=false,font=normalsize,labelfont=sf,textfont=sf]{subfig}
\usepackage{textcomp}
\usepackage{stfloats}
\usepackage{verbatim}
\usepackage{graphicx}
\usepackage{cite}
\usepackage {tablefootnote}

\usepackage{array}
\usepackage{amssymb,amsmath,latexsym}
\usepackage{amssymb}
\usepackage{amsmath}
\usepackage{extarrows}
\usepackage{makecell}
\usepackage{cancel}

\usepackage{multirow}
\usepackage{multicol}

\usepackage{url}
\usepackage{hyperref}
\hypersetup{colorlinks=true,linkcolor=blue,anchorcolor=blue,citecolor=blue}
\usepackage[ruled,linesnumbered]{algorithm2e}
\usepackage{enumerate}

\newtheorem{theorem}{Theorem}
\newtheorem{lemma}{Lemma}

\newtheorem{definition}{Definition}
\newtheorem{fact}{Fact}
\newtheorem{assumption}{Assumption}

\newcommand{\mb}{\mathbf}

\newcommand{\mc}{\mathcal}

\begin{document}

\title{Marker+Codeword+Marker: A Coding Structure for Segmented Single-Insdel/-Edit Channels}

%\author{IEEE Publication Technology,~\IEEEmembership{Staff,~IEEE,}
\author{
	\IEEEauthorblockN{Zhen Li$^\dag$, Xuan He$^\S$ and Xiaohu Tang$^\S$}\\
	\IEEEauthorblockA{$^\dag$School of Mathematics, $^\S$School of Information Science and Technology\\
		Southwest Jiaotong University\\
		Email: lz-math@my.swjtu.edu.cn;xhe@swjtu.edu.cn;xhutang@swjtu.edu.cn
	}
}
        % <-this % stops a space
%\thanks{This paper was produced by the IEEE Publication Technology Group. They are in Piscataway, NJ.}% <-this % stops a space
%\thanks{Manuscript received October 26, 2023; revised December 8, 2023.}}

%\markboth{Journal of \LaTeX\ Class Files,~Vol.~1, No.~2, December~2023}%

%\IEEEpubid{0000--0000~\copyright~2023 IEEE}

\maketitle

\begin{abstract}
An insdel refers to a deletion or an insertion, and an edit refers to an insdel or a substitution.
In this paper, we consider the segmented single-insdel (resp. single-edit) channel, where the channel's input bit stream is partitioned into segments of length $n$ and each segment can suffer from at most a single insdel (resp. edit) error.
The value of $n$ is known to the receiver but the boundaries of segments are not.
We propose to encode each segment following a marker+codeword+marker structure, where the two markers are carefully selected and the codewords are chosen from Varshamov-Tenegolts (VT) codes.
In this way, we are able to construct a new class of binary codes that can correct segmented single-insdel errors.
Our codes have the lowest redundancy of $\log_2(n-6)+7$ bits and are the first one that has linear-time encoder/decoder in the literature.
Moreover, by enhancing the VT codes and one of the markers, we are able to construct the first class of binary codes that can correct segmented single-edit errors.
This class of codes has redundancy $\log_2(n-9)+10$ bits and has linear-time encoder/decoder.
\end{abstract}


\section{Introduction}
%\IEEEPARstart{A}{n}
An insdel refers to a deletion or an insertion, and an edit refers to an insdel or a substitution.
Let $\mathbf{x} = \mathbf s^{(1)} \mathbf s^{(2)} \cdots \mathbf s^{(i)}\cdots$ represent a bit stream, where each $\mathbf s^{(i)}$ is a segment/substring of length $n$.
Suppose $\mathbf{x}$ is transmitted over a segmented single-deletion/-insdel/-edit channel, where each segment $\mathbf s^{(i)}$ can suffer from at most one deletion/insdel/edit error, respectively.
The receiver holds the knowledge of $n$, and its task is to correct the possible errors in each segment and to identify the boundaries of segments.
This model was initially introduced by Liu and Mitzenmacher \cite{DBLP:journals/tit/LiuM10}, representing a simplification of the general deletion channel where there is no limitation on the number or positions of the deletion errors.
Designing practically efficient codes with high rates and linear-time encoding/decoding algorithms for the general deletion channel presents a formidable challenge, even in basic settings where only one or two errors may occur \cite{10.1214/08-PS141,DBLP:journals/tit/BrakensiekGZ18,DBLP:conf/isit/SimaGB20, DBLP:conf/isit/SimaGB20b,DBLP:journals/tit/SimaB21,DBLP:journals/tit/HaeuplerS21,DBLP:journals/tit/SongPCH22,DBLP:journals/tit/SongC23}.
Instead, the segmentation assumption not only simplifies this design problem, but also stands to hold true with the advancements in channel quality in practical settings.

To help the reader better understand the segmented single-deletion/-insdel/-edit channel, we present three toy examples below where the channel input has three segments each of which consists of 5 bits.

\begin{itemize}
	\item \textbf{Segmented single-deletion channel:} In this scenario, only a single deletion error is permitted within each segment. For example, consider the input and output sequences:
	\[
	00111~00011~00001\rightarrow
	00111~0001\cancel1~\cancel00001,
	\]
	where $\cancel 1$ denotes an $1$-deletion and $\cancel 0$ denotes a $0$-deletion.
	
	\item \textbf{Segmented single-insdel channel:} Here, only a single insdel error is allowed within each segment. For example, consider the input and output sequences:
	\[
	00111~00011~00001\rightarrow
	0011\cancel1~0001\check11~\check000001,
	\]
	where $\check1$ and $\check0$ denote an 1-insertion and a 0-insertion, respectively.
	
	\item \textbf{Segmented single-edit channel:} In this setting, only a single edit error can occur to each segment. For example, consider the input and output sequences:
	\[
	00111~00011~00001\rightarrow
	0\underline1111~0001\underline0~\cancel00001,
	\]
	where $\underline{1}$ represents a 0-to-1 substitution and $\underline{0}$ represents an 1-to-0 substitution.
\end{itemize}

First consider the simplest case where there is only one segment and at most one insdel/edit error.
The well-known binary Varshamov-Tenegolts (VT) codes defined below perform as a class of optimal single-insdel/single-edit correcting codes \cite{Levenshtein1966BinaryCC}.
\begin{definition}\label{definition: VT code}
	Given integers $m>k>0$ and $a\in\mathbb Z_m \triangleq \{0, 1, \ldots, m-1\}$, the VT code $VT_a(m;k)$ is defined by the following congruence equation:
	\begin{equation*}\label{VT}
		VT_a(m;k)=\left\{\mathbf v\in\{0,1\}^k:\sum_{i=1}^{n}i\cdot v_i\equiv a\pmod m\right\}.
	\end{equation*}
\end{definition}

By the pigeonhole principle, there is some integer $a$ such that the code $VT_a(m;k)$ can achieve the lowest redundancy of $\log_2m$ (in this paper, we measure the redundancy of any codes in bits).
When $m=k+1$, the code can correct an insdel, and the lowest redundancy, i.e., $\log_2(k+1)$, can be achieved at $a=0$ \cite{Sloane2002}. When $m= 2k$, the code can correct an edit, and the lowest redundancy is $\log_2k+1$.
Particularly, we note that there exist efficient encoding and decoding algorithms which operate in linear time with respect to the code length $k$ \cite{DBLP:journals/tit/Ferreira98, Sloane2002, DBLP:journals/tit/Tenengolts84, DBLP:journals/tit/CaiCGKN21}.

When there are multiple segments in the input sequence, limited results are available in the field of error-correcting codes (ECCs). Liu \textit{et al.} introduced a set of sufficient conditions for searching codes that defend against segmented single-deletion/-insertion errors \cite{DBLP:journals/tit/LiuM10}.
However, as stated by the authors, the pursuit of optimal codes naturally corresponds to the challenge of identifying maximum independent sets within graphs, a task known to be NP-complete.
In \cite{DBLP:journals/tit/AbroshanVF18}, the authors constructed codes by selecting subsets of VT codes that fulfill specific yet straightforward prefix and suffix conditions.
In the binary case, the redundancy in each segment of length $n$ amounts to $r\geq\log_2(n+1)+2$ bits for segmented single-deletion case and $r\geq\log_2(n+1)+2.5$ bits for segmented single-insertion case, respectively.
Furthermore, in the context of the segmented single-insdel channel, the proposed code achieves redundancy $r\geq\log_2(n+1)+7$ bits, but there is no encoding algorithm provided.

Recent advancements in this area, as presented in \cite{DBLP:journals/tcom/JiaoLMHH22}, have enhanced the selection of subsets from binary VT codes, resulting in a reduction of code redundancy by $\log_2 1.5$ bits when dealing with the segmented single-deletion channel.
In response to the growing importance of ECCs in DNA-based data storage systems, Cai \textit{et al.}\cite{DBLP:conf/isit/CaiKMN21} introduced an encoder for quaternary segmented single-insdel ECCs, yielding a redundancy of approximately $\log_2n+6+6\log_23$ bits in each segment.
Besides, a class of quaternary segmented single-insdel ECCs with local weight constraints is presented, whose code redundancy is $\log_2n+24$ bits.
More recently, Yan \textit{et al.} \cite{DBLP:journals/tetc/YanLW23} proposed quaternary codes to combat the segmented single-edit channel by introducing markers to VT codewords, incurring a cost of $2\log_2n'+14$ bits in each segment, where $n'$ satisfies $n=n'+\lceil\log_2n'\rceil+7$.
However, this result cannot apply to the binary case since it is based on a \textit{separate function} that relies on a quaternary alphabet.
Table \ref{compare} summarizes the existing results for segmented channels.

\begin{table*}
	\renewcommand{\arraystretch}{1.3}
	\begin{center}
		\caption{Known Results for Segmented Single-Error Channels}\label{compare}
		\begin{tabular}{|c|c|c|c|}
			\hline
			Reference & Alphabet & Single-error type & Redundancy in each segment of length $n$ (in bits) \\
			\hline
			\multirow{2}{*}{\cite{DBLP:journals/tit/LiuM10}} & \multirow{7}{*}{Binary}
			& {\color{red}Deletion} & \multirow{2}{*}{/} \\
			\cline{3-3}
			&			& {\color{blue}Insertion} & \\
			\cline{1-1}
			\cline{3-4}
			\multirow{3}{*}{\cite{DBLP:journals/tit/AbroshanVF18}} &
			& {\color{red}Deletion} & $r\geq\log_2(n+1)+2$ \\
			\cline{3-4}
			& & {\color{blue}Insertion}  & $r\geq\log_2(n+1)+2.5$ \\
			\cline{3-4}
			& & Insdel  & $r\geq\log_2(n+1)+7$ \\
			\cline{1-1}
			\cline{3-4}
			\multirow{2}{*}{\cite{DBLP:journals/tcom/JiaoLMHH22}} &
			& {\color{red}Deletion} & $r\geq\log_2(n+1)+2-\log_21.5$ \\
			\cline{3-4}
			&			& {\color{blue}Insertion} & $r\geq\log_2(n+1)+2.5$\\
			\hline
			\hline
			\multirow{3}{*}{\cite{DBLP:journals/tit/AbroshanVF18}} & \multirow{5}{*}{Quaternary}
			& {\color{red}Deletion} & $r\geq\log_2n+6-2\log_23$ \\
			\cline{3-4}
			&			& {\color{blue}Insertion} & $r\geq\log_2n+8$\\
			\cline{3-4}
			&			& Insdel & $r\geq\log_2n+16$\\
			\cline{1-1}
			\cline{3-4}
			\cline{3-4}
			\cite{DBLP:conf/isit/CaiKMN21} & & Insdel & $\log_2n+6+6\log_23$\\
			\cline{1-1}
			\cline{3-4}
			\cite{DBLP:journals/tetc/YanLW23} & & {\color{cyan}Edit} & $2\log_2n'+14$, with $n=n'+\lceil\log_2n'\rceil+7$\\
			\hline
			\hline
			Theorem \ref{insdel-ecc} &  \multirow{2}{*}{Binary}  & Insdel & $\log_2(n-6)+7$ \\
			\cline{1-1}
			\cline{3-4}
			Theorem \ref{edit-ecc} &   & {\color{cyan}Edit} & $\log_2(n-9)+10$\\
			\hline
		\end{tabular}
	\end{center}
\end{table*}

%	Natural questions arise:
%	\begin{enumerate}
	%		\item Is it possible to find a low-redundancy binary codes capable of effectively correcting segmented single-insdel errors that better than \cite{DBLP:journals/tit/AbroshanVF18}, or present an efficient encoding algorithm?
	%		\item Is it possible to find low-redundancy binary codes capable of effectively correcting segmented single-edit errors?
	%	\end{enumerate}
In this paper, we consider the construction of (binary) ECCs for the segmented single-insdel/single-edit channels.
To this end, we propose a marker+codeword+marker structure to encode each segment of length $n$, where the markers are carefully selected, and the codewords are selected from the VT codes of length $k$.
Our main results and contributions are described below.
\begin{itemize}
	\item In Theorem \ref{insdel-ecc}, we construct a code $\mathcal C$ that can correct segmented single-insdel errors.
	Each segment of $\mathcal C$ follows the marker+codeword+marker structure, where the first marker has one bit, the second marker has 6 bits, and the codewords are chosen from the VT code $VT_a(k+1;k)$, resulting in a total redundancy of $\log_2(n-6)+7$ bits for each segment.
	Our code $\mathcal C$ has the lowest redundancy among the segmented single-insdel ECCs (see Table \ref{compare}) and is the first one which has linear-time encoder/decoder in the literature.
	
	\item In Theorem \ref{edit-ecc}, we further construct a code $\mathcal C'$ that can correct segmented single-edit errors.
	Each segment of $\mathcal C'$ also follows the marker+codeword+marker structure, where the first marker has 3 bits, the second marker is the same as that of $\mathcal C$, and the codewords are chosen from the VT code $VT_a(2k;k)$, leading to a total redundancy of $\log_2(n-9)+10$ bits for each segment.
	Our code $\mathcal C'$ has linear-time encoder/decoder, and to the best of our knowledge, it is the first class of binary segmented single-edit ECCs in the literature.
\end{itemize}

The subsequent sections of this paper are organized as follows.
Section \ref{sec2} introduces some notations, followed by a review of VT codes. Section \ref{sec-insdel} is dedicated to presenting the segmented single-insdel ECCs, and Section \ref{sec-edit} focuses on the construction of segmented single-edit ECCs. Finally, we draw our conclusions in Section \ref{sec5}.

\section{Preliminary}\label{sec2}
In this paper, we denote scalars by normal letters, (binary) strings/sequences/vectors by bold lowercase letters, and sets by calligraphic uppercase letters.
For any positive integer $n$, denote $[n] \triangleq \{1, 2, \ldots, n\}$, and denote $0^n$ (resp. $1^n$) as a string of $n$ zeros (resp. ones).
For any vector $\mathbf{v} = v_1 v_2\cdots v_n$ and any integers $1 \leq i \leq j \leq n$, denote $\mathbf v[i:j] \triangleq v_i v_{i+1}\cdots v_{j}$ and $\mathbf v[i:\infty] \triangleq v_i v_{i+1}\cdots v_{n}$.
For any two vectors $\mb{u}$ and $\mb{v}$, their concatenation is denoted by $\mb{u}\mb{v}$.
Their insdel distance is denoted by $d^{\textrm{insdel}}(\mb u, \mb v)$, which refers to the minimum number of insdels applying to $\mb u$ such that $\mb u$ becomes $\mb v$.
The single-insdel error ball of $\mb{u}$ is defined by $\mc{B}^{\textrm{single-insdel}}(\mb{u}) \triangleq \{\mb{v}: d^{\textrm{insdel}}(\mb u, \mb v) \leq 1\}$.
The aforementioned notations/definitions automatically apply to edits by replacing `insdel' with `edit'.

For any given VT codes $VT_a(m;k)$ and a vector $\mb{u} = u_1 u_2 \cdots u_{k}$, we call $\sum_{i=1}^{k} i\cdot u_i \pmod m$ as the VT syndrome of $\mb{u}$.
By Definition \ref{definition: VT code}, $\mb{u} \in VT_a(m;k)$ if and only if $\mb{u}$'s VT syndrome equals $a$.
%%Since the VT codes $VT_a(k+1;k)$ and $VT_a(2k;k)$ can correct an insdel and an edit, respectively,
%%the following fact holds and it will be widely used in later discussions.
The VT codes $VT_a(k+1;k)$ and $VT_a(2k;k)$ can correct an insdel and an edit, respectively.
Therefore, the minimum insdel/edit code distance, the minimum value of arbitrary two codewords of $VT_a(m;k)$, is at least 3 for both $m=k+1,2k$.
The following fact holds and it will be widely used in later discussions.
\begin{fact}\label{fact}
	For any $\mb{u} \in VT_a(m;k)$ with $m = k+1$ or $m = 2k$, denote $\mb{v}$ as any possible result by applying a deletion and an insertion to $\mb{u}$.
	Then, either $\mb{v} = \mb{u}$ or $\mb{v} \notin VT_a(m;k)$, but not both, holds.
\end{fact}

Throughout this paper, let $\mathbf{x} = \mathbf s^{(1)} \mathbf s^{(2)} \cdots \mathbf s^{(t)} = x_1x_2\cdots x_N\in\{0,1\}^N$ denote the bit stream transmitted over the segmented single-error channel.
Here for $i \in [t]$, the $i$-th segment $\mathbf s^{(i)}$ is of length $n = N/t$.
Denote the segmented single-insdel ball of $\mathbf{x}$ by $\mc{B}_{\textrm{seg}}^{\textrm{single-insdel}}(\mb{x}) \triangleq \{\mathbf u^{(1)} \mathbf u^{(2)} \cdots \mathbf u^{(t)}: \mathbf u^{(i)} \in \mc{B}^{\textrm{single-insdel}}(\mathbf s^{(i)}), i \in [t]\}$, and the segmented single-edit ball of $\mathbf{x}$ by $\mc{B}_{\textrm{seg}}^{\textrm{single-edit}}(\mb{x}) \triangleq \{\mathbf u^{(1)} \mathbf u^{(2)} \cdots \mathbf u^{(t)}: \mathbf u^{(i)} \in \mc{B}^{\textrm{single-edit}}(\mathbf s^{(i)}), i \in [t]\}$.
Denote the receiving string by $\mb{y} = y_1 y_2\cdots y_M$, and $\mb{y} \in \mc{B}_{\textrm{seg}}^{\textrm{single-insdel}}(\mb{x})$ for the segmented single-insdel channel and $\mb{y} \in \mc{B}_{\textrm{seg}}^{\textrm{single-edit}}(\mb{x})$ for the segmented single-edit channel.

\section{Coding for Segmented Single-Insdel Channel}\label{sec-insdel}

In this section, we introduce a new class of segmented single-insdel ECCs.
Let $\mathbf a_0=000010,\mathbf a_1=111101,\mathbf b_0=0$ and $\mathbf b_1=1$.
For a valid VT code $VT_a(k+1;k)$, we define a segment set $\mathcal S$ as follows:
\[
\mathcal S=\left\{\mathbf b\mathbf v \mathbf a:
\makecell{
	\mathbf v = v_1 \cdots v_k \in VT_a(k+1;k)\setminus\{0^k,1^k\}
	\\
	\mathbf a\in\{\mathbf a_0,\mathbf a_1\},\mathbf b\in\{\mathbf b_0,\mathbf b_1\}
}
\right\}.
\]
Each segment in $\mc{S}$ follows a marker+codeword+marker structure.
Based on $\mc{S}$, we construct a class of segmented single-insdel ECCs below.

\begin{theorem}\label{insdel-ecc}
	With notations as before. Define the code
	\begin{equation*}
		\mathcal C=
		\left\{
		\mathbf s^{(1)}\mathbf s^{(2)}\cdots\mathbf s^{(t)}\in\{0,1\}^{nt}:\mathbf s^{(i)}=\mathbf b^{(i)}\mathbf v^{(i)}\mathbf a^{(i)}\in\mathcal S
		\right\}
	\end{equation*}
	under the following two rules:
	\begin{enumerate}[(\text{R}1)]\label{rule}
		\item For $1\leq i\leq t$, $\mathbf b^{(i)}= \mathbf b_0$ if $v_1^{(i)}=0$, otherwise $\mathbf b^{(i)}= \mathbf b_1$;
		\item For $1\leq i\leq t-1$, $\mathbf a^{(i)}=\mathbf a_1$ if $\mathbf b^{(i+1)}=\mathbf b_0$, otherwise $\mathbf a^{(i)}=\mathbf a_0$.
		Moreover, $\mathbf a^{(t)}=\mathbf a_0$.
	\end{enumerate}
	Then $\mathcal C$ is capable of correcting segmented single-insdel errors with redundancy $\log_2(n-6)+7$ bits and with linear-time encoding/decoding algorithm.
\end{theorem}

It is clear that $\mathcal C$ has redundancy $\log_2(n-6)+7 = \log_2(k+1)+7$ and has linear-time encoding algorithm since for $i \in [t]$, each $\mathbf v^{(i)}$ in $\mc{C}$ is taken from $VT_a(k+1;k)\setminus\{0^k,1^k\}$.
To prove Theorem \ref{insdel-ecc}, it suffices to present a linear-time decoding algorithm.

Recall that $\mathbf x=x_1x_2\cdots x_N=\mathbf s^{(1)}\mathbf s^{(2)}\cdots \mathbf s^{(t)}\in\mathcal C$ represents the transmitted string, and
$\mathbf y=y_1 y_2 \cdots y_M \in\mathcal B_{\textrm{seg}}^{\textrm{single-insdel}}(\mathbf s^{(1)}\mathbf s^{(2)}\cdots\mathbf s^{(t)})$ represents the received string subject to segmented single-insdel errors.
At a high level, to successfully recover $\mathbf x$ from $\mb{y}$, our objective is to sequentially find positions/integers $p_1, p_2, \ldots, p_{t-1}$ such that:
\begin{enumerate}
	\item we can retrieve $\mathbf v^{(i)}$ from $\mathbf y[p_{i-1}+1:p_i], \forall i \in [t]$ (where we set $p_0=0$ and $p_t = M$ for convenience), and
	\item $\mathbf y[p_i+1:\infty]\in\mathcal B_{\textrm{seg}}^{\textrm{single-insdel}}(\mathbf s^{(i+1)}\mathbf s^{(i+2)}\cdots\mathbf s^{(t)}), \forall i \in [t-1]$.
\end{enumerate}

Assume that $p_1, \ldots, p_{i-1}$ have been successfully determined for some $i \in [t - 1]$, and $\mathbf y$ is replaced by $\mathbf y[p_{i-1}+1:\infty]$. (This situation is always true for $i = 1$.)
We now illustrate how to retrieve $\mathbf v^{(i)}$ and determine $p_i$.

Since at most one insdel can occur in each segment, we can judge $\mathbf a^{(i)} = \mathbf a_1 = 111101$ or $\mathbf a^{(i)} = \mathbf a_0 = 000010$ by simply applying the majority principle to $\mathbf y[k+2:k+4]$.
Moreover, the marker $\mathbf a^{(i+1)}\in\{111101,000010\}$ (if exists) can also be uniquely determined by only reading a few bits, as shown below.
\begin{lemma}\label{lemma: insdel-marker}
	Denote by $n_0,n_1$ the number of 0s and 1s in $\mathbf y[2k+9:2k+14]$ (which should originally be $\mathbf a^{(i+1)}$ if no error occurs), $n_0',n_1'$ the number of 0s and 1s in $\mathbf y[2k+9:2k+13]$, respectively. We have
	\begin{equation*}\label{insdel-determine}
		\mathbf a^{(i+1)}=
		\begin{cases}
			111101, &	n_0<n_1 \textrm{ or } (n_0=n_1 \text{ and } n_0'<n_1'),	\\
			000010, &	n_0>n_1 \textrm{ or } (n_0=n_1 \text{ and } n_0'>n_1').
		\end{cases}
	\end{equation*}
\end{lemma}
\begin{IEEEproof}
	The proof is omitted since it is straightforward to verify the correctness.
\end{IEEEproof}

As we can always determine the values of $\mathbf a^{(i)}$ and $\mathbf a^{(i+1)}$, without loss of generality (due to the symmetric property between $\mathbf a_1$ and $\mathbf a_0$ as well as that between $\mathbf b_1$ and $\mathbf b_0$), we simplify the proof using the following assumption.

\begin{assumption}\label{assumption: insdel}
	$\mathbf a^{(i)} = \mathbf a^{(i+1)} =\mathbf a_1=111101$, $\mathbf b^{(i+1)} = \mathbf b_0=0$, and $v_1^{(i+1)} = 0$.
\end{assumption}

Recall that each segment $\mathbf s^{(i)}\in\mathcal S$ takes the form $\mathbf s^{(i)}=\mathbf b^{(i)}\mathbf v^{(i)}\mathbf a^{(i)}$ for $i \in [t]$, and $\mathbf v^{(i)}=v_1^{(i)}v_2^{(i)}\cdots v_k^{(i)}\in VT_a(k+1;k)$.
In what follows, we present two useful lemmas to retrieve $\mathbf v^{(i)}$ and determine $p_i$.	
\begin{lemma}\label{insdel-retrieve}
	It suffices to use $\mathbf y[1:k+7]$ to retrieve $\mathbf v^{(i)}$.
\end{lemma}
\begin{IEEEproof}
	If $\mathbf y[2:k+1] \in VT_a(k+1;k)$, we simply have $\mathbf v^{(i)} = \mathbf y[2:k+1]$.
	If $\mathbf y[2:k+1] \notin VT_a(k+1;k)$, then $\mathbf b^{(i)} \mathbf v^{(i)}$ must suffer from an insdel and $\mathbf a^{(i)}=111101$ remains unchanged.
	Note that $\mathbf y[k+2:k+7]$ should originally be $\mathbf a^{(i)}$.
	Therefore, if $\mathbf y[k+2:k+7]$=11101* (resp. $\mathbf y[k+2:k+7]$=*11110) where * denotes an arbitrary bit, a deletion (resp. an insertion) occurs to $\mathbf b^{(i)} \mathbf v^{(i)}$, such that $\mathbf y[2:k-1] \in \mc{B}^{\textrm{single-insdel}}(\mathbf v^{(i)})$
	(resp. $\mathbf y[2:k+1] \in \mc{B}^{\textrm{single-insdel}}(\mathbf v^{(i)})$)
	can be used to retrieve $\mathbf v^{(i)}$ in linear time and $p_i = k+6$
	(resp. $p_i = k+8$)
	satisfies $\mathbf y[p_i+1:\infty]\in\mathcal B_{\textrm{seg}}^{\textrm{single-insdel}}(\mathbf s^{(i+1)}\mathbf s^{(i+2)}\cdots\mathbf s^{(t)})$.
\end{IEEEproof}

As shown above, $p_i$ can be determined if $\mathbf y[2:k+1] \notin VT_a(k+1;k)$.
If $\mathbf y[2:k+1] \in VT_a(k+1;k)$, the following result is required for determining $p_i$.

\begin{lemma}\label{insdel-p}
	Suppose $\mathbf y[2:k+1] \in VT_a(k+1;k)$.
	It suffices to use $\mathbf y[k+2:2k+14]$ to determine a $p_i$ such that $\mathbf y[p_i+1:\infty]\in\mathcal B_{\textrm{seg}}^{\textrm{single-insdel}}(\mathbf s^{(i+1)}\mathbf s^{(i+2)}\cdots\mathbf s^{(t)})$.
\end{lemma}
\begin{IEEEproof}
	We defer the proof to Appendix \ref{apdx-insdel}.
\end{IEEEproof}

Based on the above preparations, it is straightforward to prove Theorem \ref{insdel-ecc}.

\textbf{Proof of Theorem \ref{insdel-ecc}.}
We establish the theorem by presenting a decoding algorithm in Algorithm \ref{Alg-insdel}.
With the help of Lemma \ref{insdel-retrieve}, we can retrieve each $\mathbf v^{(i)},i \in [t]$ in linear time with respect to $n$.
Combining with Lemma \ref{insdel-p}, each $p_i, i \in [t-1]$ can be determined in linear time with respect to $n$ such that
$\mathbf y[p_i+1:\infty]\in\mathcal B_{\textrm{seg}}^{\textrm{single-insdel}}(\mathbf s^{(i+1)}\mathbf s^{(i+2)}\cdots\mathbf s^{(t)})$.
The proof is completed.

\begin{center}
	\begin{algorithm}[htbp]\label{Alg-insdel}
		\SetAlgoLined
		\KwIn{$\mathbf y\in
			\mathcal B_{\textrm{seg}}^{\textrm{single-insdel}}(\mathbf s^{(1)}\mathbf s^{(2)}\cdots \mathbf s^{(t)})$.}
		\KwOut{ $\mathbf v^{(i)},i=1,2,\ldots,t$.}
		\For{$i\leftarrow1$ \textrm{to} $t$}{
			\eIf{$\mathbf y[2:k+1]\notin VT_a(k+1;k)$}{Correct the error to obtain $\mathbf v^{(i)}$ and determine $p_i$ by Lemma \ref{insdel-retrieve}.}
			{$\mathbf v^{(i)}\leftarrow\mathbf y[2:k+1]$ and determine $p_i$ by Lemma \ref{insdel-p}.}
			$\mathbf y\leftarrow\mathbf y[p_i+1:\infty]$.
		}
		\caption{Segmented single-insdel channel decoder}
	\end{algorithm}
\end{center}

\section{Coding for Segmented Single-Edit Channel}\label{sec-edit}

In this section, we present the first class of segmented single-edit ECCs.
Similar to the construction of segmented single-insdel codes in Section \ref{sec-insdel}, we also encode each segment here following a marker+codeword+marker structure.
Reset the candidate markers as $\mathbf a_0=000010$, $\mathbf a_1=111101$, $\mathbf b_0=100$ and $\mathbf b_1=011$.
For a valid VT code $VT_a(2k;k)$, define a segment set
\[
\mathcal S'=\left\{\mathbf b\mathbf v \mathbf a:
\makecell{
	\mathbf v=v_1v_2\cdots v_k\in VT_a(2k;k)\setminus\{0^k,1^k\}
	\\
	\mathbf a\in\{\mathbf a_0,\mathbf a_1\},\mathbf b\in\{\mathbf b_0,\mathbf b_1\}
}
\right\}.
\]
Based on $\mc{S}'$, we construct a class of segmented single-edit ECCs below.

\begin{theorem}\label{edit-ecc}
	With notations as before. Construct the code by
	\begin{equation*}
		\begin{aligned}
			\mathcal C'=
			\left\{
			\mathbf s^{(1)}\mathbf s^{(2)}\cdots \mathbf s^{(t)}\in\{0,1\}^{nt}:\mathbf s^{(i)}=\mathbf b^{(i)}\mathbf v^{(i)}\mathbf a^{(i)}\in\mathcal S'
			\right\}
		\end{aligned}
	\end{equation*}
	under the two rules (R1) and (R2) from Theorem \ref{insdel-ecc}.
	Then $\mathcal C'$ can correct segmented single-edit errors with redundancy $\log_2(n-9)+10$ bits and with linear-time encoding/decoding algorithm.
\end{theorem}

Since each $\mathbf v^{(i)}$ in $\mc{C}'$ is chosen from $VT_a(2k;k)\setminus\{0^k,1^k\}$ for $i \in [t]$, then $\mathcal C'$ has redundancy $\log_2(n-9)+10 = \log_2k+10$ and has linear-time encoding algorithm.
To complete the proof of Theorem \ref{edit-ecc}, it suffices to present a linear-time decoding algorithm.

Similar to the proof of Theorem \ref{insdel-ecc}, let the input and output of the segmented single-edit channel be denoted by $\mathbf x=x_1x_2\cdots x_N=\mathbf s^{(1)}\mathbf s^{(2)}\cdots \mathbf s^{(t)}\in\mathcal C'$ and
$\mathbf y=y_1 y_2 \cdots y_M \in\mathcal B_{\textrm{seg}}^{\textrm{single-edit}}(\mathbf s^{(1)}\mathbf s^{(2)}\cdots\mathbf s^{(t)})$, respectively.
Our task is to illustrate how to find each $p_i, i \in [t-1]$ such that
\begin{enumerate}
	\item we can retrieve $\mathbf v^{(i)}$ from $\mathbf y[p_{i-1}+1:p_i], \forall i \in [t]$ (set $p_0=0$ and $p_t = M$ for convenience), and
	\item $\mathbf y[p_i+1:\infty]\in\mathcal B_{\textrm{seg}}^{\textrm{single-edit}}(\mathbf s^{(i+1)}\mathbf s^{(i+2)}\cdots\mathbf s^{(t)}), \forall i \in [t-1]$.
\end{enumerate}

Since each segment can suffer from at most one edit, we can judge $\mathbf a^{(i)} = \mathbf a_1 = 111101$ or $\mathbf a^{(i)} = \mathbf a_0 = 000010$ by simply applying the majority principle to $\mathbf y[k+4:k+6]$.
Similar to Lemma \ref{lemma: insdel-marker} and Assumption \ref{assumption: insdel}, we use Lemma \ref{lemma: edit-marker} to determine $\mathbf a^{(i+1)}$ and Assumption \ref{assumption: edit} to simplify the analysis, respectively.

\begin{lemma}\label{lemma: edit-marker}
	Denote by $n_0,n_1$ the number of 0s and 1s in $\mathbf y[2k+13:2k+18]$, $n_0',n_1'$ the number of 0s and 1s in $\mathbf y[2k+13:2k+17]$, respectively. We have
	\begin{equation*}\label{edit-marker}
		\mathbf a^{(i+1)}=
		\begin{cases}
			111101, & n_0<n_1,\textrm{ or } (n_0=n_1 \text{ and } n_0'<n_1'),\\
			000010, & n_0>n_1,\textrm{ or } (n_0=n_1 \text{ and } n_0'>n_1').
		\end{cases}
	\end{equation*}
\end{lemma}
\begin{IEEEproof}
	It is straightforward to verify the correctness such that the proof is omitted.
\end{IEEEproof}

\begin{assumption}\label{assumption: edit}
	$\mathbf a^{(i)} = \mathbf a^{(i+1)} =\mathbf a_1=111101$, $\mathbf b^{(i+1)} = \mathbf b_0=100$, and $v_1^{(i+1)} = 0$.
\end{assumption}

In the following, similar to Lemmas \ref{insdel-retrieve} and \ref{insdel-p}, we provide Lemmas \ref{edit-retrieve} and \ref{edit-p} to retrieve $\mathbf v^{(i)}$ and determine $p_i$, respectively.

\begin{lemma}\label{edit-retrieve}
	It suffices to use $\mathbf y[1:k+9]$ to retrieve $\mathbf v^{(i)}$.
\end{lemma}

\begin{IEEEproof}
	If $\mathbf y[4:k+3]\in VT_a(2k;k)$, we simply have $\mathbf v^{(i)}=\mathbf y[4:k+3]$.
	If $\mathbf y[4:k+3]\notin VT_a(2k;k)$, then $\mathbf b^{(i)}\mathbf v^{(i)}$ must suffer from an edit, and $\mathbf a^{(i)}=111101$ remains unchanged.
	Note that $\mathbf y[k+4:k+9]$ should originally be $\mathbf a^{(i)}=111101$.
	Therefore,
	if $\mathbf y[k+4:k+9]$=11101*, a deletion occurs to $\mathbf b^{(i)}\mathbf v^{(i)}$ such that $\mathbf y[4:k+2]\in\mathcal B^{\textrm{single-edit}}(\mathbf v^{(i)})$;
	if $\mathbf y[k+4:k+9]$=111101, a substitution occurs to $\mathbf b^{(i)}\mathbf v^{(i)}$ such that $\mathbf y[4:k+3]\in\mathcal B^{\textrm{single-edit}}(\mathbf v^{(i)})$;
	if $\mathbf y[k+4:k+9]$=*11110, an insertion occurs to $\mathbf b^{(i)}\mathbf v^{(i)}$ such that $\mathbf y[4:k+4]\in\mathcal B^{\textrm{single-edit}}(\mathbf v^{(i)})$.
	Correspondingly, $\mathbf y[4:k+2]/\mathbf y[4:k+3]/\mathbf y[4:k+4]$ can be used to retrieve $\mathbf v^{(i)}$ in linear time and $p_i=k+8/k+9/k+10$ satisfies $\mathbf y[p_i+1:\infty]\in\mathcal B_{\textrm{seg}}^{\textrm{single-edit}}(\mathbf s^{(i+1)}\mathbf s^{(i+2)}\cdots\mathbf s^{(t)})$.
\end{IEEEproof}

As shown above, $p_i$ can be determined if $\mathbf y[4:k+3]\notin VT_a(2k;k)$. If $\mathbf y[4:k+3]\in VT_a(2k;k)$, the following result is needed for determining $p_i$.
\begin{lemma}\label{edit-p}
	Suppose $\mathbf y[4:k+3] \in VT_a(2k;k)$.
	It suffices to use $\mathbf y[k+4:2k+18]$ to determine a $p_i$ such that $\mathbf y[p_i+1:\infty]\in\mathcal B_{\textrm{seg}}^{\textrm{single-edit}}(\mathbf s^{(i+1)}\mathbf s^{(i+2)}\cdots\mathbf s^{(t)})$.
\end{lemma}
\begin{IEEEproof}
	We defer the proof to Appendix \ref{apdx-edit}.
\end{IEEEproof}

Based on the above preparations, it is straightforward to prove Theorem \ref{edit-ecc}.

\textbf{Proof of Theorem \ref{edit-ecc}.}
We establish the theorem by presenting a decoding algorithm in Algorithm \ref{Alg-edit}.
Based on Lemma \ref{edit-retrieve}, we can retrieve each $\mathbf v^{(i)},i \in [t]$ in linear time with respect to $n$.
Further based on Lemma \ref{edit-p}, each $p_i, i \in [t-1]$ can be determined in linear time with respect to $n$ such that
$\mathbf y[p_i+1:\infty]\in\mathcal B_{\textrm{seg}}^{\textrm{single-edit}}(\mathbf s^{(i+1)}\mathbf s^{(i+2)}\cdots\mathbf s^{(t)})$.
The proof is completed.

\begin{center}
	\begin{algorithm}[htbp]\label{Alg-edit}
		\SetAlgoLined
		\KwIn{$\mathbf y\in
			\mathcal B_{\textrm{seg}}^{\textrm{single-edit}}(\mathbf s^{(1)}\mathbf s^{(2)}\cdots \mathbf s^{(t)})$.}
		\KwOut{ $\mathbf v^{(i)},i=1,2,\ldots,t$. }
		\For{$i \leftarrow 1$ to $t$}{
			\eIf{$\mathbf y[4:k+3]\notin VT_a(2k;k)$}{Correct the error to obtain $\mathbf v^{(i)}$ and determine $p_i$ by Lemma \ref{edit-retrieve}.}
			{$\mathbf v^{(i)}\leftarrow\mathbf y[4:k+3]$ and determine $p_i$ by Lemma \ref{edit-p}.}
			$\mathbf y\leftarrow\mathbf y[p_i+1:\infty]$.
		}
		\caption{Segmented single-edit channel decoder}
	\end{algorithm}
\end{center}

\section{Conclusion}\label{sec5}
In this paper, we proposed a marker+codeword+marker structure to construct ECCs for segmented single-insdel/single-edit channels.
By using carefully selected markers and codewords from the VT code $VT_a(k+1;k)$, we constructed a new class of binary ECCs for segmented single-insdel channels.
Our codes have the lowest redundancy and are the first one which can be encoded and decoded in linear time.
Moreover, by using longer markers and enhanced codewords from $VT_a(2k;k)$, we constructed the first class of binary ECCs for segmented single-edit channels.
This class of codes can also be encoded and decoded in linear time.
Our future research aims to prove the optimality of the codes in terms of redundancy or to reduce the redundancy.

%\section*{Acknowledgment}

{\appendices

\section{Proof of the Lemma \ref{insdel-p}}\label{apdx-insdel}

	We demonstrate this lemma through a case-by-case analysis.
First, we partition the analysis into 6 cases based on the values of the received vector $\mathbf y[k+2:k+7]$, as detailed in Table \ref{AllCases-insdel}.
Then, we identify the error types that occur to $\mathbf y[k+2:k+7]$ in each case so as to obtain $p_i$.
To this end, we may make use of $\mathbf y[2k+9:2k+14]$, which should originally be $\mathbf a^{(i+1)} = 111101$ when no error occurs to $\mathbf x$.
If this is not sufficient, we further make use of $\mathbf y[k+9:2k+8]$, which should originally be $\mathbf v^{(i+1)}$, a valid VT codeword, if no error occurs to $\mathbf x$.

	\begin{table}[ht]
		\begin{center}
			\caption{Candidates for $\mathbf y[k+2:k+7]$}\label{AllCases-insdel}
			\begin{tabular}{c|c}
				\hline
				Case					&	$\mathbf y[k+2:k+7]$	\\
				\hline
				\multirow{3}{*}{A1)}	&	011***	\\
										&	101***	\\
										&	110***	\\
				\hline
				A2)						&	1110**	\\
				\hline
				A3)						&	111100	\\
				\hline
				A4)						&	111101	\\
				\hline
				A5)						&	111110	\\
				\hline
				A6)						&	111111\\
				\hline
				\multicolumn{2}{l}{$\blacktriangle$ Here `*' denotes an arbitrary entry}\\
			\end{tabular}
		\end{center}
	\end{table}
	
	\begin{enumerate}[{A}1)]
		\item Assume there exists a `0' in $\mathbf y[k+2:k+4]$. Observing that these positions should originally be identical 1s, the error must be a 0-insertion, indicating $p_i=k+8$.
		
		\item $\mathbf y[k+2:k+5]=1110$.
		This could result from either a 0-insertion or an 1-deletion.
		In this case, $y_{k+6}=1$ always holds.
		We then partition it into the following two subcases based on the values of $y_{k+7}$.
		\item[A2.1)] $y_{k+7}=1$ implies $\mathbf y[k+2:k+7]=111011$.
    This results from only one error pattern of ${\color{red}\mathbf a^{(i)}}{\color{cyan}\mathbf b^{(i+1)}}$: {\color{red}111\cancel101}{\color{cyan}$\check1$0}\footnote{When enumerating the positions where an insdel occurs, if multiple positions lead to the same result, we only enumerate one of the positions to simplify the discussion.
    This situation usually happens when an insdel occurs to a run, and we generally choose the last one of the possible positions.
    For example in Case A2.1), the error patterns of ${\color{red}\mathbf a^{(i)}}{\color{cyan}\mathbf b^{(i+1)}}$ that make $\mathbf y[k+2:k+7]=111010$ could be {\color{red}\cancel111101}{\color{cyan}$\check1$0}, {\color{red}1\cancel11101}{\color{cyan}$\check1$0}, {\color{red}11\cancel1101}{\color{cyan}$\check1$0}, or {\color{red}111\cancel101}{\color{cyan}$\check1$0}, where the last error pattern is chosen in the discussion. Moreover, when describing an error pattern of ${\color{red}\mathbf a^{(i)}}{\color{cyan}\mathbf b^{(i+1)}}$ (or ${\color{red}\mathbf a^{(i)}}{\color{cyan}\mathbf b^{(i+1)}v_1^{(i+1)}}$), for the convenience of the reader, we color bits from ${\color{red}\mathbf a^{(i)}}$ and ${\color{cyan}\mathbf b^{(i+1)}v_1^{(i+1)}}$ in \textcolor{red}{red} and \textcolor{cyan}{cyan}, respectively.}.
    Hence we obtain $p_i=k+6$.
		
		\item[A2.2)] $y_{k+7}=0$ implies $\mathbf y[k+2:k+7]=111010$.
This results from either a 0-insertion or an 1-deletion to ${\color{red}\mathbf a^{(i)}}{\color{cyan}\mathbf b^{(i+1)}}$: {\color{red}111$\check0$101} or {\color{red}111\cancel101}{\color{cyan}0}.
		Hence more information is needed for determining the situation.
		We list all the possible error patterns for $\mathbf a^{(i)}\mathbf b^{(i+1)}v_1^{(i+1)}$ in Table \ref{2.2tab}.
		\begin{table}
			\renewcommand{\arraystretch}{1.25}
			\caption{Error Patterns for
				${\color{red}\mathbf a^{(i)}}{\color{cyan}\mathbf b^{(i+1)}v_1^{(i+1)}}$ in Case A2.2)}\label{2.2tab}
			\begin{center}
				\begin{tabular}{|c|l|c|c|}
					\hline
					Row	& Error pattern	& $\mathbf y[k+2:k+11]$	&	$p_i$ \\
					\hline
					1	&	{\color{red}111\cancel101}{\color{cyan}0\cancel0}	&	{\color{blue}111010}****	&	\multirow{4}{*}{$k+6$}	\\
					\cline{1-3}
					2	&	{\color{red}111\cancel101}{\color{cyan}00}	&	{\color{blue}111010}0***	&		\\
					\cline{1-3}
					3	&	{\color{red}111\cancel101}{\color{cyan}00$\check0$}	&	{\color{blue}111010}00**	&		\\
					\cline{1-3}
					4	&	{\color{red}111\cancel101}{\color{cyan}00$\check1$}	&	{\color{blue}111010}01**	&		\\
					\hline
					5	&	{\color{red}111\cancel101}{\color{cyan}0$\check1$0}	&	{\color{blue}111010}10**	&		\\
					\cline{1-3}
					6	&	{\color{red}111$\check0$101}{\color{cyan}0\cancel0}	&	{\color{blue}111010}10**	&		\multirow{6}{*}{$k+8$}	\\
					\cline{1-3}
					7	&	{\color{red}111$\check0$101}{\color{cyan}00}	&	{\color{blue}111010}100*	&			\\
					\cline{1-3}
					8	&	{\color{red}111$\check0$101}{\color{cyan}00$\check0$}	&	{\color{blue}111010}1000	&			\\
					\cline{1-3}
					9	&	{\color{red}111$\check0$101}{\color{cyan}00$\check1$}	&	{\color{blue}111010}1001	&		\\
					\cline{1-3}
					10	&	{\color{red}111$\check0$101}{\color{cyan}0$\check1$0}	&	{\color{blue}111010}1010	&		\\
					\cline{1-3}
					11	&	{\color{red}111$\check0$101}{\color{cyan}$\check1$00}	&	{\color{blue}111010}1100	&		\\
					\hline
				\end{tabular}
			\end{center}
		\end{table}
		
		Note from Table \ref{2.2tab} that $y_{k+8}=0$ appears only when an 1-deletion occurring to $\mathbf a^{(i)}$, i.e., 111\cancel101, hence $p_i=k+6$ immediately follows for $y_{k+8}=0$.
		For the scenario $y_{k+8}=1$, it appears in Rows 5-to-11 ($p_i=k+8$) and may appear in Row 1 ($p_i=k+6$).
		Particularly, for Row 5, both $p_i=k+6$ and $p_i=k+8$ are valid.
Therefore, the left thing is to distinguish Row 1 from Rows 6-to-11 under the assumption $y_{k+8}=1$.

		To this end, we make use of $\mathbf y[2k+9:2k+14]$, which should originally be $\mathbf a^{(i+1)} = 111101$ when no error occurs to $\mathbf x$.
In Table \ref{2.2tab2}, we enumerate the candidates for $\mathbf y[2k+9:2k+14]$ based on the error types that possibly occur to $\mathbf a^{(i)}\mathbf b^{(i+1)}$ in a general case.
For Row 1 in Table \ref{2.2tab}, we have $\mathbf y[2k+9:2k+14] \in \mathcal P$(1101**), which is shown in Table \ref{2.2tab2}.
For Rows 6-to-11 in Table \ref{2.2tab}, we have $\mathbf y[2k+9:2k+14] \in \mathcal P$($v_k^{(i+1)}$11110), which is shown in Table \ref{2.2tab2.3}.
Since $\mathcal P$($v_k^{(i+1)}11110) \cap \mathcal P$(1101**) = $\emptyset$, we can thus distinguish Row 1 from Rows 6-to-11.
That is, if $\mathbf y[2k+9:2k+14] \in \mathcal P$(1101**), set $p_i = k+6$; otherwise, set $p_i = k+8$.

		\begin{table}[htb]
			\caption{Candidates for $\mathbf y[2k+9:2k+14]$}\label{2.2tab2}
			\begin{center}
				\begin{tabular}{|c|c|c|}
					\hline
					Row	&	Error types occur to $\mathbf a^{(i)}\mathbf b^{(i+1)}$	&	$\mathbf y[2k+9:2k+14]$	\\
					\hline
					1	&	no error	& $\mathcal P$(111101) \\
					\hline
					2	&	one deletion	& $\mathcal P$(11101*)	\\
					\hline
					3	&	two deletions	& $\mc{P}$(1101**) = $\{1101^{\dag}, 110100, 110101, 110110\}$ \\
					\hline
					4	&	one insertion	&  $\mathcal P$($v_k^{(i+1)}$11110) \\
					\hline
					5	&	two insertions	& $v_{k-1}^{(i+1)}v_k^{(i+1)}$1111 \\
					\hline
					6 & one deletion+one insertion & 111101\\
					\hline
\multicolumn{3}{p{9cm}}{$\blacktriangle$ Here $\mathcal P(\cdot)$ refers to a set of possible values of $\mathbf y[2k+9:2k+14]$. More details are shown in Tables \ref{2.2tab2.2}, \ref{2.2tab2.1}, and \ref{2.2tab2.3}.}\\
\multicolumn{3}{p{9cm}}{$^{\dag}$ If $\mb{s}^{(i+2)}$ does not exist, then $y_{2k+13}$ and $y_{2k+14}$ do not exist.}\\
				\end{tabular}
			\end{center}
		\end{table}

		\begin{table}[h]
			\renewcommand{\arraystretch}{1.25}
			\caption{Possible Values of $\mathbf y[2k+9:2k+14]$ from $\mathcal P$(111101)}\label{2.2tab2.2}
			\begin{center}
				\begin{tabular}{|c|c|c|}
					\hline
					Row	&	Error types occur to $\mathbf s^{(i+1)}$	&	$\mathbf y[2k+9:2k+14](y_{2k+15})$	\\
					\hline
					1 &	no error	&	111101	\\
					\cline{1-3}
					2&	insertion occurs to $\mathbf v^{(i+1)}$	&	$v_k^{(i+1)}11110$	\\
					\cline{1-3}
					3&	deletion occurs to $\mathbf v^{(i+1)}$	&	11101*	\\
					\cline{1-3}
					\multirow{3}{*}{4}&	\multirow{3}{*}{0-insertion to $\mathbf a^{(i+1)}$}	&	$\check011110(1),1\check01110(1)$,	\\
					&	&	$11\check0110(1),111\check010(1)$,	\\
					&	&	$11110\check0(1),111101(\check0)$\\
					\cline{1-3}
					5&	1-insertion to $\mathbf a^{(i+1)}$	&	$1111\check10(1),111101(\check1)$	\\
					\cline{1-3}
					6&	0-deletion to $\mathbf a^{(i+1)}$	&	1111\cancel01*	\\
					\cline{1-3}
					7&	1-deletion to $\mathbf a^{(i+1)}$	&	111\cancel101*,11110\cancel1*\\
					\hline
				\end{tabular}
			\end{center}
		\end{table}		
		
		\begin{table}[tb]
			\renewcommand{\arraystretch}{1.25}
			\caption{Possible Values of $\mathbf y[2k+9:2k+14]$ from $\mathcal P$(11101*)}\label{2.2tab2.1}
			\begin{center}
				\begin{tabular}{|c|c|c|}
					\hline
					Row	&	Error types occur to $\mathbf s^{(i+1)}$	&	$\mathbf y[2k+9:2k+14]$	\\
					\hline
					1 &	no error	&	11101*\\
					\cline{1-3}
					2&	insertion occurs to $\mathbf v^{(i+1)}$	&	111101	\\
					\cline{1-3}
					3&	deletion occurs to $\mathbf v^{(i+1)}$	&	1101**	\\
					\cline{1-3}
					\multirow{2}{*}{4}&	\multirow{2}{*}{0-insertion to $\mathbf a^{(i+1)}$}	&	$\check011101,1\check01101,11\check0101$	\\
					&	&	$1110\check01,11101\check0$	\\
					\cline{1-3}
					5&	1-insertion to $\mathbf a^{(i+1)}$	&	$111\check101,11101\check1$	\\
					\cline{1-3}
					6&	0-deletion to $\mathbf a^{(i+1)}$	&	111\cancel01**	\\
					\cline{1-3}
					7&	1-deletion to $\mathbf a^{(i+1)}$	&	11\cancel101**,1110\cancel1**	\\
					\hline
				\end{tabular}
			\end{center}
		\end{table}

		\begin{table*}[htb]
			\renewcommand{\arraystretch}{1.25}
			\caption{Possible Values of $\mathbf y[2k+9:2k+14]$ from $\mathcal P(v_k^{(i+1)}$11110)}\label{2.2tab2.3}
			\begin{center}
				\begin{tabular}{|c|c|c|}
					\hline
					Row	&	Error types occur to $\mathbf s^{(i+1)}$	&	$\mathbf y[2k+9:2k+14](y_{2k+15})$	\\
					\hline
					1 &	$\textrm{no error}$	&	$v_k^{(i+1)}$11110\\
					\cline{1-3}
					2&	insertion occurs to $\mathbf v^{(i+1)}$	&	$v_{k-1}^{(i+1)}v_k^{(i+1)}1111,\check*v_k^{(i+1)}$1111	\\
					\cline{1-3}
					3&	deletion occurs to $\mathbf v^{(i+1)}$	&	111101	\\
					\cline{1-3}
					\multirow{2}{*}{4}&	\multirow{2}{*}{0-insertion to $\mathbf a^{(i+1)}$}	&	
					$v_k^{(i+1)}\check01111,v_k^{(i+1)}1\check0111,v_k^{(i+1)}11\check0$11	\\
					&	&	$v_k^{(i+1)}111\check01,v_k^{(i+1)}11110(\check0)$	\\
					\cline{1-3}
					5&	1-insertion to $\mathbf a^{(i+1)}$	&	$v_k^{(i+1)}1111\check1,v_k^{(i+1)}11110(\check1)$	\\
					\cline{1-3}
					6&	0-deletion to $\mathbf a^{(i+1)}$	&	$v_k^{(i+1)}1111\cancel01$	\\
					\cline{1-3}
					7&	1-deletion to $\mathbf a^{(i+1)}$	&	$v_k^{(i+1)}111\cancel101,v_k^{(i+1)}11110(\cancel1)$\\
					\hline
				\end{tabular}
			\end{center}
		\end{table*}

		\item Assume that $\mathbf y[k+2:k+7]=111100$.
In this case, the error pattern of $\mb{a}^{(i)}$ is either {\color{red}11110$\check0$1} or {\color{red}11110\cancel1}.
We can thus determine $p_i$ similarly as in Case A2.2), and we omit the details to save space.
		
		\item Now $\mathbf y[k+2:k+7]=111101$.
		We note that when $y_{k+8}=1$, we can simply let $p_i=k+8$ since $\mathbf y[k+2:k+8]=1111011$, indicating that $y_{k+8}$ must be an inserted bit.
		In what follows, we only consider the case $y_{k+8}=0$ and list the candidates of $\mathbf y[k+8:k+11]$ in Table \ref{4tab}.
		\begin{table}[ht]
			\renewcommand{\arraystretch}{1.25}
			\caption{Error Patterns for
				${\color{red}\mathbf a^{(i)}}{\color{cyan}\mathbf b^{(i+1)}v_1^{(i+1)}}$ for Case A4)}\label{4tab}
			\begin{center}
				\begin{tabular}{|c|l|c|c|}
					\hline
					Row	& Error pattern	& $\mathbf y[k+2:k+11]$	&	$p_i$ \\
					\hline
					1	&	{\color{red}111101}{\color{cyan}0\cancel0}	&	{\color{blue}111101}0***	&	\multirow{5}{*}{$k+7$}	\\
					\cline{1-3}
					2	&	{\color{red}111101}{\color{cyan}00}	&	{\color{blue}111101}00**	&		\\
					\cline{1-3}
					3	&	{\color{red}111101$\check0$}{\color{cyan}0\cancel0}	&	{\color{blue}111101}00**	&		\\
					\cline{1-3}
					4	&	{\color{red}11110\cancel1}{\color{cyan}$\check1$00}	&	{\color{blue}111101}00**	&		\\
					\cline{1-3}
					5	&	{\color{red}111101}{\color{cyan}00$\check0$}	&	{\color{blue}111101}000*	&		\\
					\hline
					6	&	{\color{red}111101$\check0$}{\color{cyan}00}	&	{\color{blue}111101}000*	&	\multirow{3}{*}{$k+8$}	\\
					\cline{1-3}
					7	&	{\color{red}111101$\check0$}{\color{cyan}00$\check0$}	&	{\color{blue}111101}0000	&	\\
					\cline{1-3}
					8	&	{\color{red}111101$\check0$}{\color{cyan}00$\check1$}	&	{\color{blue}111101}0001	&			\\
					\hline
					9	&	{\color{red}111101}{\color{cyan}00$\check1$}	&	{\color{blue}111101}001*	&	$k+7$	\\
					\hline
					10	&	{\color{red}111101$\check0$}{\color{cyan}0$\check1$0}	&	{\color{blue}111101}0010	&	$k+8$	\\
					\hline
					11	&	{\color{red}111101}{\color{cyan}0$\check1$0}	&	{\color{blue}111101}010*	&	$k+7$	\\
					\hline
					12	&	{\color{red}111101$\check0$}{\color{cyan}$\check1$00}	&	{\color{blue}111101}0100	&	$k+8$	\\
					\hline
				\end{tabular}
			\end{center}
		\end{table}

		To identify which type of error occurs, we need to read $\mathbf y[2k+9:2k+14]$, which should originally be $\mathbf a^{(i+1)}$ in $\mathbf s^{(i)}\mathbf s^{(i+1)}$.
		Moreover, there is no more error in $\mathbf s^{(i+1)}$ for all the cases except when no error occurs to $\mathbf b^{(i+1)}v_1^{(i+1)}$, which are the scenarios of Rows 2 and 6 in Table \ref{4tab}.
		No error occurs to $\mathbf a^{(i)}\mathbf b^{(i+1)}v_1^{(i+1)}$ in the Row 2 in Table \ref{4tab}, hence $\mathbf y[2k+9:2k+14] \in \mathcal P(111101)$, which is shown in Table \ref{2.2tab2.2};
		a 0-insertion occurs to $\mathbf a^{(i)}\mathbf b^{(i+1)}v_1^{(i+1)}$ in the Row 6 in Table \ref{4tab}, hence $\mathbf y[2k+9:2k+14] \in \mathcal P(v_k^{(i+1)}$11110), which is shown in Table \ref{2.2tab2.3}.

		We note that when receiving $y_{k+9}=1$, it must be one of the scenarios shown in Rows 1, 11 and 12, indicating that $\mathbf y[2k+9:2k+14]$ is 11101*, $v_k^{(i+1)}$11110 and $v_{k-1}^{(i+1)}v_k^{(i+1)}$1111, respectively.
		These three candidates can be completely distinguished by the number and position of 0s.
		Thus the corresponding error pattern can be correspondingly determined, and $p_i$ follows.

		When receiving $y_{k+9}=0$, it can appear in Rows 1-to-10.
		We need to read $y_{k+10}$, and partition it into two subcases based on the values of $y_{k+10}$.
		\item[A4.1)]
		If $y_{k+10}=1$, it must be one of the scenarios shown in Rows 1-to-4, 9 and 10 in Table \ref{4tab}.
		Specifically, we have a total of 5 situations as given below:\\
		(\textbf{a}) First consider the scenario in Row 2.
		No error occurs to $\mathbf a^{(i)}\mathbf b^{(i+1)}$.
		There may be an error occurs to $\mathbf v^{(i+1)}\mathbf a^{(i+1)}$, making $\mathbf y[2k+9:2k+14] \in \mathcal P(111101)$, which is shown in \ref{2.2tab2.2}.
		
		(\textbf{b}) For the Row 1, a 0-deletion occurs to $\mathbf s^{(i+1)}$, hence no further error can occur.
		We have $\mathbf y[2k+9:2k+14]=$11101*.
		By the Rows 3 and 4 in Table \ref{2.2tab2.2}, there exist two error patterns 11101* and 111$\check0$10 which lead to $\mathbf y[2k+9:2k+14]=$11101*$\in\mathcal P(111101)$.
		However, we can ignore this confusable scenario and simply let $p_i=k+7$ in this case since the errors to $\mathbf a^{(i)}$ are the same for both situations.
		
		(\textbf{c}) For the Rows 3 and 4, we can regard the error the same as Row 2, hence $p_i=k+7$ follows.
		
		(\textbf{d}) For the Row 9, an 1-insertion occurs to $\mathbf s^{(i+1)}$, hence no further error can occur.
		We have $\mathbf y[2k+9:2k+14]=v_k^{(k+1)}$11110.
		By the Rows 2 and 6 in Table \ref{2.2tab2.2}, there exist two error patterns $v_k^{(i+1)}$11110 and 1111\cancel01* which make $v_k^{(k+1)}11110\in\mathcal P(111101)$.
		However, we can ignore this confusable scenario and simply let $p_i=k+7$ in this case since the errors to $\mathbf a^{(i)}$ are the same for both situations.
		
		(\textbf{e}) For the Row 10, there are two insertions occurring, hence no further error can occur to $\mathbf s^{(i+1)}$.
		Now we have $\mathbf y[2k+9:2k+14]=v_{k-1}^{(i+1)}v_k^{(i+1)}1111$.
		According to the Row 6 in Table \ref{2.2tab2.2}, a 0-deletion occurs to $\mathbf a^{(i+1)}:$ 1111\cancel01*, implying a possibility that $v_{k-1}^{(i+1)}v_k^{(i+1)}\in\mathcal P$(111101).
		Consequently, we assert that $v_{k-1}^{(i+1)}v_k^{(i+1)}=11$ and the symbol `*' in 11111* equals to 1, which implies there must be an 1-insertion occurs to the left to $\mathbf s^{(i+2)}$.
		Under this assertion, we obtain that the two bits following $v_{k-1}^{(i+1)}v_k^{(i+1)}1111$ are $y_{2k+15}y_{2k+16}=01$, the last two bits of $\mathbf a^{(i+1)}$, since there are two insertions occurring to segments $\mathbf s^{(i)}\mathbf s^{(i+1)}$.
		For situation \textbf{(a)}, $y_{2k+15}y_{2k+16}=00$ corresponds to $\mathbf b^{(i+2)}v_1^{(i+2)}$ since there is one deletion along with one insertion occurring to segments $\mathbf s^{(i)}\mathbf s^{(i+1)}$.
		This leads to a contradiction. Hence we can obviously determine the error in segment $\mathbf s^{(i)}$, then $p_i$ follows.

		Combining all of these 5 cases, each error type can be certainly distinguished, and then $p_i$ can be correspondingly determined.
		
		\item[A4.2)]
		If $y_{k+10}=0$, it appears in Rows 1-to-8 in Table \ref{4tab}.
		We have 4 situations in total:
		
		(\textbf{a'}) For the Row 2, this is already presented in the Case A4.1) \textbf{(a)} (see also in Table \ref{2.2tab2.2}). There is no error to $\mathbf a^{(i)}\mathbf b^{(i+1)}\mathbf v^{(i+1)}$, and $\mathbf y[2k+9:2k+14] \in \mathcal P(111101)$.
		
		(\textbf{b'}) For the Row 1, we can simply let $p_i=k+7$ in this case since the errors to $\mathbf a^{(i)}$ are the same for both situations, as already shown in Case A4.1) \textbf{(b)}.
		Now we have that 0-deletion occurs to $\mathbf s^{(i+1)}$ and $\mathbf y[2k+9:2k+14]$=11101*.
		
		(\textbf{c'}) For the Rows 3 and 4, we can regard the error the same as Row 2, hence $p_i =k+7$ follows.
		
		(\textbf{d'}) For the Rows 5 and 6, we can always regard as that a 0-insertion occurs to $\mathbf a^{(i)}$ and no error occurs to $\mathbf b^{(i+1)}v_1^{(i+1)}$.
		Therefore, $\mathbf y[2k+9:2k+14]$ lies in $\mathcal P(v_k^{(i+1)}$11110), and all the possible received results are listed in Table \ref{2.2tab2.3}.
		We note that, many error patterns to $\mathbf v^{(i+1)}$ shall lead to $\mathbf y[2k+9:2k+14]$ also belongs to $\mathcal P$(11101*), $\mathcal P(111101)$, which incurs confusion when we identify the error pattern to $\mathbf a^{(i)}\mathbf b^{(i+1)}\mathbf v^{(i+1)}$.
		Therefore, we need distinguish them and correspondingly determine a $p_i$ for each case.
%		This leads to several cases, and some of them may be confusable with other situations because of the intersections between the sets $\mathcal P$(11101*), $\mathcal P(111101)$ and $\mathcal P(v_k^{(i+1)}11110)$.
		
		To this end, we highlight our approach: first, judge which situation the received $\mathbf y[2k+9:2k+14]$ corresponds to, and then check the VT codeword $\mathbf v^{(i+1)}$ in segment $\mathbf s^{(i+1)}$ to help determine the error pattern that occurred to $\mathbf a^{(i)}\mathbf b^{(i+1)}\mathbf v^{(i+1)}$.
		Finally, we can determine a suitable $p_i$.

		We partition the values of $\mathbf y[2k+9:2k+14]$ in Table \ref{2.2tab2.3} into the following 9 cases: \textbf{(d'-1)}-to-\textbf{(d'-9)}, and analyze them one by one.
		
		(\textbf{d'-1}) $\mathbf y[2k+9:2k+14]=$*11110.
		We note there are 3 error patterns for situation \textbf{(a')} such that *11110$\in \mc{P}(111101)$.
		Specifically, the error types are as follows:
		\romannumeral1) an insertion occurs to $\mathbf v^{(i+1)}$,
		\romannumeral2) an 1-insertion occurs to $\mathbf a^{(i+1)}:1111\check10$,
		\romannumeral3) a 0-deletion occurs to $\mathbf a^{(i+1)}$1111\cancel01*.
		
		Recall that $\mathbf y[k+9:2k+8]$ should originally be $\mathbf v^{(i+1)}$ in $\mathbf s^{(i)}\mathbf s^{(i+1)}$ if no error occurs.
		Therefore, the vector $\mathbf y[k+10:2k+9]$ in $\mathbf s^{(i)}\mathbf s^{(i+1)}$ can be regarded as a result of a VT codeword that suffers an insertion along with one deletion, which would not be a VT codeword unless the insertion offset the deletion.
		We note that there is no error occurring to $\mathbf v^{(i+1)}$ for situation \textbf{(d')}, and now $\mathbf v^{(i+1)}$ has shifted one bit to the right since a 0-insertion occurred to $\mathbf a^{(i)}$.

		We now make use of $\mathbf y[k+10:2k+9]$.
		If this vector does not form a VT codeword, then it must be the situation \textbf{(a')}, hence $p_i=k+7$ follows.
		If it indeed forms a VT codeword, then it could either be situation \textbf{(a')} via a 0-insertion occurring left to $v_2^{(i+1)}\cdots v_k^{(i+1)}$, or situation \textbf{(d')}.
		By noting that $\mathbf b^{(i+1)}v_1^{(i+1)}=00$, then for situation \textbf{(a')}, the 0-insertion occurring left to $v_2^{(i+1)}\cdots v_k^{(i+1)}$ can be regarded as a 0-insertion right to the last bit of segment $\mathbf s^{(i)}$.
		Thus we can simply let $p_i=k+8$ such that $\mathbf y[p_i+1:\infty]\in\mathcal B_{\textrm{seg}}^{\textrm{single-insdel}}(\mathbf s^{(i+1)}\mathbf s^{(i+2)}\cdots\mathbf s^{(t)})$ satisfied for both situations \textbf{(a')} and \textbf{(d')}.

		(\textbf{d'-2}) $\mathbf y[2k+9:2k+14]=$*11101. It could also be situation \textbf{(a')} since *11101$\in\mathcal P(111101)$, as predicted in the Rows 4 and 7 in Table \ref{2.2tab2.2}: 11110$\check0$, 111101($\check0$) and 11110\cancel1*.
		If `*' is 0, then it must be situation \textbf{(d')}; we then only consider that `*' is 1.
		In this case, $\mathbf y[2k+9:2k+14]=$111101 could be situation \textbf{(a')} or \textbf{(d')}.
		We note that the error in segment $\mathbf s^{(i+1)}$ would locate in $\mathbf a^{(i+1)}$ for situation \textbf{(a')}, hence the VT codeword remains unchanged.

		There are three error patterns for situation \textbf{(d')}:
		\romannumeral1) a 0-insertion occurs to $\mathbf a^{(i+1)}: v_k^{(i+1)}111\check01$ and $v_k^{(i+1)}=1$,
		\romannumeral2) a deletion occurs to $\mathbf v^{(i+1)}$, which results in $v_2^{(i+1)}\cdots \cancel{v_j^{(i+1)}}v_{j+1}^{(i+1)}\cdots v_k^{(i+1)}$,
		\romannumeral3) an 1-deletion occurs to $\mathbf a^{(i+1)}:v_k^{(i+1)}111\cancel101$ and $v_k^{(i+1)}=1$.
		For these three cases, we can list the corresponding bits of $\mathbf y[k+9:2k+8]$, which should originally be a VT codeword:
		\romannumeral1) $0v_1^{(i+1)}\cdots v_{k-1}^{(i+1)}$,
		\romannumeral2) $0v_1^{(i+1)}\cdots \cancel{v_j^{(i+1)}}v_{j+1}^{(i+1)}\cdots v_k^{(i+1)}$,
		\romannumeral3) $0v_1^{(i+1)}\cdots v_{k-1}^{(i+1)}$.
		Each one of them is a result of a VT codeword via an insertion and one deletion, which cannot be a VT codeword unless it remains unchanged.

		Based on the analyses above, we assert that $p_i$ can be determined by computing the VT syndrome of $\mathbf y[k+9:2k+8]$.
		Specifically, if the syndrome changes, then it must be the situation \textbf{(d')}, and $p_i=k+8$;
		if the syndrome remains unchanged, then we have $p_i=k+7$.

		(\textbf{d'-3}) $\mathbf y[2k+9:2k+14]=$*11011.
			It could also be situations \textbf{(a')} and \textbf{(b')} since *11011$\in\mathcal P(111101)$, as predicted in Rows 3 and 7 in Table \ref{2.2tab2.2}: 11101*, 111\cancel101*.
		If `*' is 0, then it must be situation \textbf{(d')}; therefore, we only consider the symbol `*' is 1 in what follows.

		For situation \textbf{(a')}, no error occurs to $\mathbf a^{(i)}\mathbf b^{(i+1)}v_1^{(i+1)}$ and $\mathbf y[2k+9:2k+14]$ lies in $\mathcal P(111101)$.
		To satisfy the assumption, *11011$\in\mathcal P(111101)$ should hold.
		Consequently, the error could be: (1)11101$\check1$(00)\footnote{Here the notation ($\cdot$) refers to the following bits. This shall be useful in identifying the error patterns.} or 111\cancel101$\check1$(00).
		That is, there must be a deletion (either occurring to $\mathbf v^{(i+1)}$ or $\mathbf a^{(i+1)}$) along with an 1-insertion occurring to the left of $\mathbf b^{(i+2)}$, which implies that $y_{2k+15}y_{2k+16}=00$.
		For situation \textbf{(b')}, `11101*=*11011' indicates that $\mathbf y[2k+9:2k+14]=111011$.
		Recall that there is a 0-deletion occurs to $\mathbf s^{(i+1)}$. Thus, $y_{2k+14}=1$ belongs to the segment $\mathbf s^{(i+2)}$, which is an inserted bit, implying that $y_{2k+15}y_{2k+16}=00$.
		For situation \textbf{(d')}, one insertion occurs, and a 0 is inserted to $\mathbf a^{(i+1)}$, giving $y_{2k+15}y_{2k+16}=01$.
		Therefore, they can be distinguished by reading $y_{2k+15}y_{2k+16}$.
		Note that $p_i$ is the same for \textbf{(a')} and \textbf{(b')}. Then, $p_i$ for \textbf{(d')} can be determined.
		
		(\textbf{d'-4}) $\mathbf y[2k+9:2k+14]\in$\{*10111,*01111,0*1111\}.
		It must be the situation \textbf{(d')}, leading to $p_i=k+8$.
		
		(\textbf{d'-5}) $\mathbf y[2k+9:2k+14]=$111101. It could also be situation \textbf{(a')}.
		For situation \textbf{(a')}, 111101$\in\mathcal P(111101)$ should be satisfied.
		Thus the error cannot occur to $\mathbf v^{(i+1)}$, as detailed in Table \ref{2.2tab2.2}, hence $\mathbf y[k+9:2k+8]$ remains unchanged, still a VT codeword.
		For situation \textbf{(d')}, a 0-insertion occurs to $\mathbf a^{(i)}$, and $\mathbf y[2k+9:2k+14]$ lies in $\mathcal P(v_k^{(i+1)}$11110). Thus the error patterns are of two types:
		\romannumeral1) a deletion to $\mathbf v^{(i+1)}$,
		\romannumeral2) an 1-deletion occurs to $\mathbf a^{(i+1)}$: $v_k^{(i+1)}111\cancel101$ along with $v_k^{(i+1)}=1$,
		\romannumeral3) a 0-insertion occurs: $v_k^{(i+1)}111\check0$1 along with $v_k^{(i+1)}=1$.
		Consequently, $\mathbf y[k+9:2k+8]$ is resulted from a VT codeword via a 0-insertion and a deletion for \romannumeral1), hence it has a possibility of being a VT codeword; $\mathbf y[k+9:2k+8]$ results from a VT codeword via a 0-insertion and an 1-deletion for \romannumeral2), hence it cannot be a VT codeword.
		
		We can then make use of $\mathbf y[k+9:2k+8]$ to determine $p_i$ for these events.
		Specifically, when it forms a VT codeword, then the situation could be \textbf{(a')}, or \textbf{(d')} under the error type \romannumeral1).
		For situation \textbf{(d')}, we note that the 0-insertion and the deletion to $\mathbf v^{(i+1)}1111$ does not change the VT codeword by the Fact \ref{fact}.
		Therefore, $p_i=k+7$ holds as long as the VT syndrome of $\mathbf y[k+9:2k+8]$ remains unchanged, since now the errors for situations \textbf{(a')} and \textbf{(d')} are exactly the same in the segment $\mathbf s^{(i)}$;
		otherwise, we assert it must be the situation \textbf{(d')}, hence $p_i=k+8$ follows.
		
		(\textbf{d'-6}) $\mathbf y[2k+9:2k+14]=$*11110. It could also be situation \textbf{(a')} as predicted in Rows 2, 4 and 5 in Table \ref{2.2tab2.2}: an insertion occurring to $\mathbf v^{(i+1)}$, or an insertion occurring to $\mathbf a^{(i+1)}$, which is $\check0$11110(1) or 1111$\check1$0(1).
		For situation \textbf{(d')}, the error cannot occur to $\mathbf v^{(i+1)}$, as predicted in Table \ref{2.2tab2.3}, hence $\mathbf y[k+10:2k+9]$ is still a VT codeword.
		Thus we can make use of received $\mathbf y[k+9:2k+8]$ (it should be a VT codeword if no error occurs) at first.

		We have the following conclusion: if $\mathbf y[k+9:2k+8]$ forms a VT codeword, then it must be the situation \textbf{(a')}, hence $p_i=k+7$ follows.
		In fact, if it is the situation \textbf{(d')}, then $\mathbf y[k+9:2k+8]=0v_1^{(i+1)}\cdots v_{k-1}^{(i+1)}$, which is still a VT codeword. By noting that $\mathbf v^{(i+1)}=v_1^{(i+1)}\cdots v_k^{(i+1)}$, we assert that it should be $0^k$, which is impossible.

		On the other hand, if $\mathbf y[k+9:2k+8]$ does not form a VT codeword, we claim that there must be an insertion occurring to VT codeword $\mathbf v^{(i+1)}$, and we then read $y_{2k+9}$.
		If $\mathbf y[k+10:2k+9]$ cannot form a VT codeword, then it must be the situation \textbf{(a')}, hence $p_i=k+7$ follows;
		if $\mathbf y[k+10:2k+9]$ indeed forms a VT codeword, then it could either be situation \textbf{(d')}, or situation \textbf{(a')} with a 0-insertion left to $v_1^{(i+1)}$ since now $y_{k+8}y_{k+9}y_{k+10}=000$.
		We can simply let $p_i=k+8$ for this case, and $\mathbf y[p_i+1:\infty]\in\mathcal B_{\textrm{seg}}^{\textrm{single-insdel}}(\mathbf s^{(i+1)}\mathbf s^{(i+2)}\cdots\mathbf s^{(t)})$ holds.
		
		(\textbf{d'-7}) When $\mathbf y[2k+9:2k+14]=$1*1111, it must be the situation \textbf{(d')} if the symbol `*' is 0.
		The other case, i.e., symbol `*' is 1, is left for discussion in (\textbf{e'}).
		
		(\textbf{d'-8}) $\mathbf y[2k+9:2k+14]=$*1110*. It could also be situation \textbf{(a')}, as predicted in Rows 1, 4, 5 and 7 in Table \ref{2.2tab2.2}: 111101, 11110$\check0$(1), 111101($\check0$), 111101($\check1$) and 11110\cancel1*.
		We note that the error cannot occur to $\mathbf v^{(i+1)}$, hence the VT codeword remains unchanged.
		For situation \textbf{(d')}, the error could either be a deletion or an insertion, as predicted in Rows 3/6 and 4 in Table \ref{2.2tab2.3}: 111101/$v_k^{(i+1)}$111\cancel101, or $v_k^{(i+1)}111\check01$.
		
		We make use of the received vector $\mathbf y[k+9:2k+8]$ for situation \textbf{(d')}:
		\romannumeral1) $\mathbf y[k+9:2k+8]=0v_1^{(i+1)}\cdots v_{k-1}^{(i+1)}$ if the deletion/insertion occurs to $\mathbf a^{(i+1)}$.
		Moreover, it cannot be a VT codeword since we exclude $0^k$;
		\romannumeral2) $\mathbf y[k+9:2k+8]=0v_1^{(i+1)}\cdots v_{j-1}^{(i+1)}\cancel{v_j^{(i+1)}}v_{j+1}^{(i+1)}\cdots v_k^{(i+1)}$ if the deletion occurs to $\mathbf v^{(i+1)}$.
		Moreover, it cannot form a VT codeword unless the deleted bit is 0 and locate in the same run as the inserted 0.
		
		Therefore, we can distinguish situations \textbf{(a')} and \textbf{(d')} via computing the VT syndrome of received vector $\mathbf y[k+9:2k+8]$.
		Specifically, when $\mathbf y[k+9:2k+8]$ is not a VT codeword, then it is the situation \textbf{(d')}, hence $p_i=k+8$ follows;
		otherwise, it could either be situation \textbf{(a')}, or situation \textbf{(d')} via the error \romannumeral2).
		However, we can simply let $p_i=k+7$ since the error to \textbf{(d')} can be regarded as no error, which is the same as \textbf{(a')}.
		
		(\textbf{d'-9}) For the case $\mathbf y[2k+9:2k+14]=$*11111, we leave it for discussion in situation (\textbf{e'}).

		(\textbf{e'}) For the Rows 4 and 5, we have $\mathbf y[2k+9:2k+14]=$**1111, hence it could also be situations \textbf{(a')} and \textbf{(d')}.
		Specifically, for the situation \textbf{(a')}, it is of the error pattern: 1111\cancel01*(00) with `*' being $\check1$ , presented in Row 6 in Table \ref{2.2tab2.2}; for the situation \textbf{(d')}, there are four error patterns: $v_{k-1}^{(i+1)}v_k^{(i+1)}1111,\check*v_k^{(i+1)}1111,v_k^{(i+1)}1111\check1$, and $v_k^{(i+1)}$1111\cancel01 with $v_k^{(i+1)}=1$, presented in Rows 2, 5 and 6 in the Table \ref{2.2tab2.3}, respectively.
		
		Consequently, $y_{2k+15}y_{2k+16}=00$ for situation \textbf{(a')},
		$y_{2k+15}y_{2k+16}\in\{01,00,10\}$ for situation \textbf{(d')},
		and $y_{2k+15}y_{2k+16}=01$ for situation \textbf{(e')}.
		Therefore, when $y_{2k+15}y_{2k+16}\neq00$, we can simply let $p_i=k+8$ since there is an insertion occurring to $\mathbf a^{(i)}$ for both situations \textbf{(d')} and \textbf{(e')}.
		When $y_{2k+15}y_{2k+16}=00$, it could either be the situation \textbf{(a')}, or situation \textbf{(d')} with 0-deletion: $v_k^{(i+1)}$1111\cancel01.
		Under this condition, the error cannot occur to $\mathbf v^{(i+1)}$ for both situations \textbf{(a')} and \textbf{(d')}.
		Consequently, one can distinguish them by identifying whether $\mathbf y[k+9:2k+8]$ is a VT codeword as previously shown, and $p_i$ can be obtained for both of these situations.
		Here we omit the routine details.

		\item $\mathbf y[k+2:k+7]=111110$.
In this case, the error pattern of $\mb{a}^{(i)}$ is either {\color{red}1111$\check1$01} or {\color{red}1111\cancel01}.
We can thus determine $p_i$ similarly as in Case A2.2), and we omit the details to save space.

		\item $\mathbf y[k+2:k+7]=111111$.
		It can result from only one error pattern of ${\color{red}\mathbf a^{(i)}}{\color{cyan}\mathbf b^{(i+1)}v_1^{(i+1)}}$: {\color{red}1111\cancel01}{\color{cyan}$\check1$00}.
		Hence $p_i=k+6$ follows.
		
	\end{enumerate}
	
	In summary, for each case in Table \ref{AllCases-insdel}, we can obtain $p_i$ such that 
	$\mathbf y[p_i+1:\infty]\in\mathcal B_{\textrm{seg}}^{\textrm{single-insdel}}(\mathbf s^{(i+1)}\mathbf s^{(i+2)}\cdots\mathbf s^{(t)})$.
This completes the proof.

\section{Proof of the Lemma \ref{edit-p}}\label{apdx-edit}

	Similar to the proof of Lemma \ref{insdel-p}, we demonstrate this lemma through a case-by-case analysis.
	Here, we partition the values of $\mathbf y[k+4:k+11]$ into 18 cases, as detailed in Table \ref{AllCases-edit}.
	We analyze each case individually.

	\begin{table}[ht]
		\begin{center}
			\caption{Candidates for $\mathbf y[k+4:k+11]$}\label{AllCases-edit}
			\begin{tabular}{c|ccc|ccc|c}
				\cline{1-2}
				\cline{4-5}
				\cline{7-8}
				Case	&	$\mathbf y[k+4:k+11]$	&&	Case	&	$\mathbf y[k+4:k+11]$	&&	Case	&	$\mathbf y[k+4:k+11]$	\\
				\cline{1-2}
				\cline{4-5}
				\cline{7-8}
				\multirow{3}{*}{B1)}	&	0111****	&&	B6)	&	11110011	&&	B13)	&	11111010	\\
															\cline{4-5}
															\cline{7-8}
										&	1011****	&&	B7)	&	11110100	&&	B14)	&	11111011	\\
															\cline{4-5}
															\cline{7-8}
										&	1101****	&&	B8)	&	11110101	&&	B15)	&	11111100	\\
				\cline{1-2}
				\cline{4-5}
				\cline{7-8}
				B2)						&	1110****	&&	B9)	&	11110110	&&	B16)	&	11111101	\\
				\cline{1-2}
				\cline{4-5}
				\cline{7-8}
				B3)						&	11110000	&&	B10)	&	11110111	&&	B17)	&	11111110	\\
				\cline{1-2}
				\cline{4-5}
				\cline{7-8}
				B4)						&	11110001	&&	B11)	&	11111000	&&	B18)	&	11111111	\\
				\cline{1-2}
				\cline{4-5}
				\cline{7-8}
				B5)						&	11110010	&&	B12)	&	11111001	&		\\
				\cline{1-2}
				\cline{4-5}
				\multicolumn{8}{l}{$\blacktriangle$ Here `*' denotes an arbitrary entry}\\
			\end{tabular}
		\end{center}
	\end{table}

	\begin{enumerate}[{B}1)]
		\item There is a `0' in $\mathbf y[k+4:k+6]$.
		Noting that $\mathbf y[k+4:k+8]=11110$ if no error occurs to segment $\mathbf s^{(i)}$.
		If $y_{k+8}=0$, the `0' in $\mathbf y[k+4:k+6]$ comes from substitution, and we let $p_i=k+9$.
		Otherwise, the `0' in $\mathbf y[k+4:k+6]$ is an insertion, and $p_i=k+10$.
		
		\item $\mathbf y[k+4:k+7]=1110$. This leads to three error patterns for $\mathbf a^{(i)}$
		\romannumeral1) an 1-to-0 substitution: $111\underline001$,
		\romannumeral2) a 0-insertion: $111\check0101$,
		\romannumeral3) an 1-deletion: 111\cancel101.
		We then check whether `$y_{k+8}=0$' is true.
		If it is true, error \romannumeral1) has occurred, consequently $p_i=k+9$.
		Otherwise, i.e., $y_{k+8}=1$, then an insdel has occurred.
		Noting that each segment has at most one error, it turns out that error \romannumeral3) has occurred when $y_{k+9}$ is `1'.
		Therefore we have $p_i=k+8$.
Subsequently, we consider $y_{k+9}=0$, which implies that $\mathbf y[k+4:k+9]=111010$.
We partition the analysis into four subcases based on the values of $y_{k+10}y_{k+11}$.
		\item[B2.1)] If $y_{k+10}y_{k+11}=00$, then $\mathbf y[k+4:k+11]=11101000$.
		It results from only one error pattern of {\color{red}$\mathbf a^{(i)}$}{\color{cyan}$\mathbf b^{(i+1)}v_1^{(i+1)}$}: {\color{red}111\cancel101}{\color{cyan}\cancel1000}.
		Consequently, we have $p_i=k+8$.
		
		\item[B2.2)] $y_{k+10}y_{k+11}=01$ is impossible.
		Otherwise, we would have $\mathbf y[k+4:k+11]=11101001$, which cannot result from ${\color{red}\mathbf a^{(i)}}{\color{cyan}\mathbf b^{(i+1)}v_1^{(i+1)}}=\color{red}{111101}\color{cyan}{1000}$.
		
		\item[B2.3)] If $y_{k+10}y_{k+11}=10$, then $\mathbf y[k+4:k+11]=11101010$.
		There are three error patterns for {\color{red}$\mathbf a^{(i)}$}{\color{cyan}$\mathbf b^{(i+1)}v_1^{(i+1)}$}: {\color{red}111\cancel101}{\color{cyan}$\check0$1000}, {\color{red}111$\check0$101}{\color{cyan}$\underline0$000}, {\color{red}111$\check0$101}{\color{cyan}\cancel1000}.
We simply let $p_i=k+10$.

		\item[B2.4)] If $y_{k+10}y_{k+11}=11$, then $\mathbf y[k+4:k+11]=11101011$.
		It results from only one error pattern of {\color{red}$\mathbf a^{(i)}$}{\color{cyan}$\mathbf b^{(i+1)}v_1^{(i+1)}$}:		{\color{red}111$\check0$101}{\color{cyan}1000}.
		Hence we have $p_i=k+10$.		
		
		\item $\mathbf y[k+4:k+11]=11110000$. 
This could result from four error patterns of {\color{red}$\mathbf a^{(i)}$}{\color{cyan}$\mathbf b^{(i+1)}v_1^{(i+1)}$}: {\color{red}11110\cancel1}{\color{cyan}\cancel1000}, {\color{red}11110\cancel1}{\color{cyan}$\underline0$000}, {\color{red}11110\underline0}{\color{cyan}\cancel1000}, {\color{red}11110\underline0}{\color{cyan}\underline0000}.
We then make use of $\mathbf y[2k+13:2k+18]$, which should originally be $\mathbf a^{(i+1)}=111101$ if no error occurs to $\mb{x}$.
Specifically, if $\mathbf y[2k+13:2k+18] = 1101\text{**}$, two deletions occur to {\color{red}$\mathbf a^{(i)}$}{\color{cyan}$\mathbf b^{(i+1)}v_1^{(i+1)}$} such that the error pattern is {\color{red}11110\cancel1}{\color{cyan}\cancel1000}, leading to $p_i = k+8$.
If $\mathbf y[2k+13:2k+18] = 11101\text{*}$, one deletion occurs to {\color{red}$\mathbf a^{(i)}$}{\color{cyan}$\mathbf b^{(i+1)}v_1^{(i+1)}$} such that the error pattern is {\color{red}11110\cancel1}{\color{cyan}$\underline0$000} or {\color{red}11110\underline0}{\color{cyan}\cancel1000}, leading to $p_i = k+8$.
Finally, if $\mathbf y[2k+13:2k+18] = 111101$, no deletion occurs to {\color{red}$\mathbf a^{(i)}$}{\color{cyan}$\mathbf b^{(i+1)}v_1^{(i+1)}$} such that the error pattern is {\color{red}11110\underline0}{\color{cyan}\underline0000}, leading to $p_i = k+9$.
	
		\item $\mathbf y[k+4:k+11]=11110001$.
		In this case, there is only one error pattern for {\color{red}$\mathbf a^{(i)}$}{\color{cyan}$\mathbf b^{(i+1)}v_1^{(i+1)}$}:
		{\color{red}11110\underline0}{\color{cyan}$\check0$1000}.
		Hence we have $p_i=k+9$.
		
		\item $\mathbf y[k+4:k+11]=11110010$.
		 We partition the analysis into four subcases based on the values of $y_{k+12}y_{k+13}$.
		
		\item[B5.1)] If $y_{12}y_{13}=00$, then $\mathbf y[k+4:k+13] = 1111001000$.
		This results from four error patterns of {\color{red}$\mathbf a^{(i)}$}{\color{cyan}$\mathbf b^{(i+1)}v_1^{(i+1)}$}:
		{\color{red}11110\underline0}{\color{cyan}1000},
		{\color{red}11110\cancel1}{\color{cyan}$\check0$1000},
		{\color{red}11110$\check0$1}{\color{cyan}\cancel1000},
		{\color{red}11110$\check0$1}{\color{cyan}\underline000(0)}.
		Hence, we can simply set $p_i=k+9$.
		
		\item[B5.2)] If $y_{12}y_{13}=01$, then $\mathbf y[k+4:k+13] = 1111001001$. 
		This results from three error patterns of 
		{\color{red}$\mathbf a^{(i)}$}{\color{cyan}$\mathbf b^{(i+1)}v_1^{(i+1)}$}:		{\color{red}11110\underline0}{\color{cyan}100\cancel0(1)},
{\color{red}11110\underline0}{\color{cyan}100\underline1},
{\color{red}11110\underline0}{\color{cyan}100$\check1$0}.
		Therefore, we can simply let $p_i=k+9$.

		\item[B5.3)] If $y_{12}y_{13}=10$, then $\mathbf y[k+4:k+13] = 1111001010$. This results from three error patterns of
		{\color{red}$\mathbf a^{(i)}$}{\color{cyan}$\mathbf b^{(i+1)}v_1^{(i+1)}$}:
		{\color{red}11110\underline0}{\color{cyan}10\underline10},	{\color{red}11110\underline0}{\color{cyan}10$\check1$00},	{\color{red}11110$\check0$1}{\color{cyan}$\check0$1000}.
We can make use of $\mathbf y[2k+13:2k+18]$ to determine $p_i$ similarly as in Case B3).

		\item[B5.4)] It is evident that $y_{12}y_{13}\neq11$.

		\item $\mathbf y[k+4:k+11]=11110011$.
 We partition the analysis into two subcases based on the values of $y_{k+12}$.
		\item [B6.1)] $y_{k+12}=0$ gives $\mathbf y[k+4:k+12]=111100110$.
		It results from three error patterns of
		{\color{red}$\mathbf a^{(i)}$}{\color{cyan}$\mathbf b^{(i+1)}v_1^{(i+1)}$}: {\color{red}11110$\check0$1}{\color{cyan}10**}, {\color{red}11110\underline0}{\color{cyan}1$\check1$000},	{\color{red}11110\underline0}{\color{cyan}1\underline100}.		We can simply let $p_i=k+10$.

		\item [B6.2)] $y_{k+12}=1$ gives $\mathbf y[k+4:k+12]=111100111$.
		It results from two error patterns of {\color{red}$\mathbf a^{(i)}$}{\color{cyan}$\mathbf b^{(i+1)}v_1^{(i+1)}$}:		{\color{red}11110$\check0$1}{\color{cyan}1\underline100},
{\color{red}11110$\check0$1}{\color{cyan}1$\check1$000}.
		Hence we have $p_i=k+10$.
		
		\item $\mathbf y[k+4:k+11]=11110100$.
		We list all possible error patterns in Table \ref{edit-7tab}.
		\begin{table}[ht]
			\renewcommand{\arraystretch}{1.25}
			\caption{Error Patterns for
				${\color{red}\mathbf a^{(i)}}{\color{cyan}\mathbf b^{(i+1)}v_1^{(i+1)}}$ for Case B7)}\label{edit-7tab}
			\begin{center}
				\begin{tabular}{|c|l|c|c|}
					\hline
					Row	& Error pattern	& $\mathbf y[k+4:k+15]$	&	$p_i$ \\
					\hline
					1	&	{\color{red}11110\cancel1}{\color{cyan}100\cancel0}	&	{\color{blue}11110100}****	&	\multirow{6}{*}{$k+8$}	\\
					\cline{1-3}
					2	&	{\color{red}11110\cancel1}{\color{cyan}1000}	&	{\color{blue}11110100}0***	&		\\
					\cline{1-3}
					3	&	{\color{red}111101}{\color{cyan}\cancel1000}	&	{\color{blue}11110100}0***	&		\\
					\cline{1-3}
					4	&	{\color{red}11110\cancel1}{\color{cyan}1000$\check0$}	&	{\color{blue}11110100}00**	&		\\
					\cline{1-3}
					5	&	{\color{red}111101}{\color{cyan}\underline0000}	&	{\color{blue}11110100}00**	&		\\
					\cline{1-3}
					6	&	{\color{red}111101$\check0$}{\color{cyan}\cancel1000}	&	{\color{blue}11110100}00**	&		\\
					\hline
					7	&	{\color{red}111101$\check0$}{\color{cyan}\underline0000}	&	{\color{blue}11110100}000*	&	$k+10$	\\
					\hline
					8	&	{\color{red}11110\cancel1}{\color{cyan}1000$\check1$}	&	{\color{blue}11110100}01**	&	\multirow{3}{*}{$k+8$}	\\
					\cline{1-3}
					9	&	{\color{red}11110\cancel1}{\color{cyan}100\underline1}	&	{\color{blue}11110100}1***	&	\\
					\cline{1-3}
					10	&	{\color{red}11110\cancel1}{\color{cyan}100$\check1$0}	&	{\color{blue}11110100}10**	&	\\
					\hline
					11	&	{\color{red}111101$\check0$}{\color{cyan}$\check0$1000}	&	{\color{blue}11110100}1000	&	$k+10$	\\
					\hline
				\end{tabular}
			\end{center}
		\end{table}
		It can be concluded that no further error occurs to segment $\mathbf s^{(i+1)}$, except for the row 2.
%		To identify the type of error that occurred, we need to determine $\mathbf a^{(i+1)}$ at first.
%		Thanks to the Lemma \ref{lemma: edit-marker}, $\mathbf a^{(i+1)}$ can be settled.
%		We then distinguish the error that occurred.
%		Without of loss of generality, we assume that $\mathbf a^{(i+1)}=111101$.
		Let us consider two cases depending on whether $y_{k+12}$ is 1.
		\item[B7.1)] $y_{k+12}=1$ appears in the Row 1 and the Rows 9-to-11.
		We can find out which type of error occurs by computing the shift of `0' in $\mathbf a^{(i+1)}=$111101.
		Specifically, a two-bits-left shift for the Row 1, a two-bits-right shift for the Row 11, and the others cannot shift two-bits.
		Consequently, $p_i$ can be determined, which is shown in the Table \ref{edit-7tab}.
		
\item[B7.2)] $y_{k+12}=0$ appears in the Rows 1-to-8.
		We first make use of $\mathbf y[2k+13:2k+18]$.
		In detail, we have $\mathbf y[2k+13:2k+18]=1101$** for Row 1;
		$\mathbf y[2k+13:2k+18]\in\mathcal P'$(11101*) for Row 2, for which $\mathcal P'(\cdot)$ refers to the possible results, and are listed in the following Table \ref{edit-7.2tab};
		\begin{table*}[t!]
			\renewcommand{\arraystretch}{1.25}
			\caption{Possible received results for $\mathbf y[2k+13:2k+18]$}\label{edit-7.2tab}
			\begin{center}
				\begin{tabular}{|c|c|c|}
					\hline
					Row	&	Error patterns for $\mathbf s^{(i+1)}$	&	$\mathbf y[2k+13:2k+18]$	\\
					\hline
					1 &	No error	&	11101*\\
					\cline{1-3}
					2&	insertion occurs to $\mathbf v^{(i+1)}$	&	111101	\\
					\cline{1-3}
					3&	deletion occurs to $\mathbf v^{(i+1)}$	&	1101**	\\
					\cline{1-3}
					4&	substitution occurs to $\mathbf v^{(i+1)}$	&	11101*	\\
					\cline{1-3}
					\multirow{2}{*}{5}&	\multirow{2}{*}{0-insertion to $\mathbf a^{(i+1)}$}
					&	$\check011101,1\check01101,11\check0101,$	\\
					&	&	$1110\check01,11101\check0$	\\
					
					\cline{1-3}
					6&	1-insertion to $\mathbf a^{(i+1)}$	&	$111\check101,11101\check1$	\\
					\cline{1-3}
					7&	0-deletion to $\mathbf a^{(i+1)}$	&	111\cancel01**	\\
					\cline{1-3}
					8&	1-deletion to $\mathbf a^{(i+1)}$	&	11\cancel101**,1110\cancel1**	\\
					\cline{1-3}
					9&	0-to-1 substitution to $\mathbf a^{(i+1)}$	&	111\underline11*	\\
					\cline{1-3}
					10&	1-to-0 substitution to $\mathbf a^{(i+1)}$	&	\underline01101*,1\underline0101*,11\underline001*,1110\underline0*	\\
					\hline
				\end{tabular}
			\end{center}
		\end{table*}
		$\mathbf y[2k+13:2k+18]=11101$* for Row 3;
		$\mathbf y[2k+13:2k+18]$=111101 for Rows 4-to-6 and 8;
		$\mathbf y[2k+13:2k+18]=$*11110 for Row 7.
		
		We can simply let $p_i=k+8$ when $\mathbf y[2k+13:2k+18]=1101$**, since there must be two deletions and one of them must occur to $\mathbf a^{(i)}$;
		we can simply let $p_i=k+8$ when $\mathbf y[2k+13:2k+18]=11101$*, since this can happen only for Rows 2 and 3 in Table \ref{edit-7tab}, and both of their error patterns can be regard as there is an 1-deletion to $\mathbf a^{(i)}$;
		we can simply let $p_i=k+8$ when $\mathbf y[2k+13:2k+18]=111101$ since this can happen only for Rows 2, 4-to-6 and 8 in Table \ref{edit-7tab}, and both of their error patterns can be regard as there is an 1-deletion to $\mathbf a^{(i)}$.
		We can simply let $p_i=k+8$ when $\mathbf y[2k+13:2k+18]$ is neither one of the above 3 patterns nor *11110, since this can happen only for Row 2 in Table \ref{edit-7tab}.
		The only thing needs to argue is when $\mathbf y[2k+13:2k+18]=$*11110.
		As shown in Rows 7 and 9 in Table \ref{edit-7.2tab}, there is a possibility of $\mathbf y[2k+13:2k+18]=$*11110$\in\mathcal P'$(11101*).
		We have, if `*=0', then it must be the Row 7, which gives $p_i=k+10$.
		We then consider `*=1', which is the only confusable case of Rows 2 and 7 in Table \ref{edit-7tab}.

		For the Row 2, we note that the error in segment $\mathbf s^{(i+1)}$ occurs to $\mathbf a^{(i+1)}$, as shown in Rows 7 and 9 in Table \ref{edit-7.2tab}.
		Thus the received vector $\mathbf y[k+12:2k+11]$ is the original VT codeword since a deletion occurred to $\mathbf a^{(i)}$, resulting in the VT codeword $\mathbf v^{(i+1)}$ moves one bit to the left.
		For the Row 7, we note that now $\mathbf y[2k+13:2k+18]=v_k^{(i+1)}$11110 and $\mathbf y[k+14:2k+13]$ is the original VT codeword, since an insertion occurred to $\mathbf a^{(i)}$ and an error occurred to $\mathbf a^{(i+1)}$, which results the VT codeword moves one bit to the right.
		To distinguish these two cases, we compute the VT syndrome of $\mathbf y[k+13:2k+12]$.
		For the case Row 2, we can write this received vector as $\mathbf u=v_2^{(i+1)}\cdots v_k^{(i+1)}1$;
		for the case Row 2, the received vector is $\mathbf u'=0v_1^{(i+1)}\cdots v_{k-1}^{(i+1)}$.
		Consequently, we have
		\begin{equation*}
			\begin{aligned}
				&\sum_{j=1}^kj\cdot u_j-\sum_{j=1}^kj\cdot v_j^{(i+1)}\\
				\equiv& -\sum_{j=1}^k v_j^{(i+1)}+k\cdot1\\
				\equiv& -wt_H(\mathbf v^{(i+1)})+k\pmod{2k},\\
				&\sum_{j=1}^kj\cdot u_j'-\sum_{j=1}^kj\cdot v_j^{(i+1)}\\
				\equiv& \sum_{j=1}^{k-1} v_j^{(i+1)}-k\cdot v_k^{(i+1)}\\
				\equiv& wt_H(\mathbf v^{(i+1)})+k-1\pmod{2k},
			\end{aligned}
		\end{equation*}
		where $wt_H(\cdot)$ refers to the hamming weight of a vector.
		These two least non-negative remainders are bounded by $k$ since $0< wt_H(\mathbf v^{(i+1)})<k$, thus they cannot reach each other's value range.
		Consequently, we can distinguish the Rows 2 and 7, and $p_i$ follows.
		It can be routinely verified that the assertion still holds for the case $\mathbf a^{(i+1)}=000010$.
		
		\item $\mathbf y[k+4:k+11]=11110101$. It results from three error patterns of
		{\color{red}$\mathbf a^{(i)}$}{\color{cyan}$\mathbf b^{(i+1)}v_1^{(i+1)}$}: {\color{red}11110\cancel1}{\color{cyan}10$\check1$00}, {\color{red}111101$\check0$}{\color{cyan}1000}, {\color{red}111101}{\color{cyan}$\check0$1000}.
		Therefore, we can simply let $p_i=k+10$.
		
		\item $\mathbf y[k+4:k+11]=11110110$. 
We partition the analysis into two subcases based on the values of $y_{k+12}$.
		
		\item[B9.1)] If $y_{k+12} = 1$, then $\mathbf y[k+4:k+12]=111101101$. 
It results from three error patterns of {\color{red}$\mathbf a^{(i)}$}{\color{cyan}$\mathbf b^{(i+1)}v_1^{(i+1)}$}: {\color{red}111101}{\color{cyan}10\underline10}, {\color{red}111101}{\color{cyan}10$\check1$00}, {\color{red}111101$\check1$}{\color{cyan}$\check0$1000}.
We can make use of $\mathbf y[2k+13:2k+18]$ to determine $p_i$ similarly as in Case B3).

		\item[B9.2)] If $y_{k+12}=0$, then $\mathbf y[k+4:k+12]=111101100$.
		We can simply let $p_i=k+9$.
		
		\item $\mathbf y[k+4:k+11]=11110111$. We list all possible error patterns in Table \ref{edit-10tab}.
As can be seen from Table \ref{edit-10tab}, we can simply set $p_i=k+10$.
		\begin{table}[h]
			\renewcommand{\arraystretch}{1.25}
			\caption{Error Patterns for
				${\color{red}\mathbf a^{(i)}}{\color{cyan}\mathbf b^{(i+1)}v_1^{(i+1)}}$ for Case B10)}\label{edit-10tab}
			\begin{center}
				\begin{tabular}{|c|l|c|c|}
					\hline
					Row	& Error pattern	& $\mathbf y[k+4:k+15]$	&	$p_i$ \\
					\hline
					1	&	{\color{red}111101$\check1$}{\color{cyan}100\cancel0}	&	{\color{blue}11110111}00**	&	\multirow{12}{*}{$k+10$}	\\
					\cline{1-3}
					2	&	{\color{red}111101}{\color{cyan}1\underline100}	&	{\color{blue}11110111}00**	&		\\
					\cline{1-3}
					3	&	{\color{red}111101$\check1$}{\color{cyan}1000}	&	{\color{blue}11110111}000*	&		\\
					\cline{1-3}
					4	&	{\color{red}111101}{\color{cyan}1$\check1$000}	&	{\color{blue}11110111}000*	&		\\
					\cline{1-3}
					5	&	{\color{red}111101$\check1$}{\color{cyan}1000$\check0$}	&	{\color{blue}11110111}0000	&		\\
					\cline{1-3}
					6	&	{\color{red}111101$\check1$}{\color{cyan}1000$\check1$}	&	{\color{blue}11110111}0001	&		\\
					\cline{1-3}
					7	&	{\color{red}111101$\check1$}{\color{cyan}100\underline1}	&	{\color{blue}11110111}001*	&	\\
					\cline{1-3}
					8	&	{\color{red}111101$\check1$}{\color{cyan}100$\check1$0}	&	{\color{blue}11110111}0010	&	\\
					\cline{1-3}
					9	&	{\color{red}111101$\check1$}{\color{cyan}10\underline10}	&	{\color{blue}11110111}010*	&	\\
					\cline{1-3}
					10	&	{\color{red}111101$\check1$}{\color{cyan}10$\check1$00}	&	{\color{blue}11110111}0100	&	\\
					\cline{1-3}
					11	&	{\color{red}111101$\check1$}{\color{cyan}1\underline100}	&	{\color{blue}11110111}100*	&	\\
					\cline{1-3}
					12	&	{\color{red}111101$\check1$}{\color{cyan}1$\check1$000}	&	{\color{blue}11110111}1000	&	\\
					\hline
				\end{tabular}
			\end{center}
		\end{table}

		\item $\mathbf y[k+4:k+11]=11111000$. 
		It results from two error patterns of 
		{\color{red}$\mathbf a^{(i)}$}{\color{cyan}$\mathbf b^{(i+1)}v_1^{(i+1)}$}: {\color{red}1111\cancel01}{\color{cyan}\underline1000},
{\color{red}1111\cancel01}{\color{cyan}\cancel1000}.
We immediately have $p_i=k+8$.

		\item $\mb y[k+4:k+11]=11111001$ is impossible.
		
		\item $\mathbf y[k+4:k+11]=11111010$. 
		It results from four error patterns of
		{\color{red}$\mathbf a^{(i)}$}{\color{cyan}$\mathbf b^{(i+1)}v_1^{(i+1)}$}:		{\color{red}1111\cancel01}{\color{cyan}$\check0$1000},
{\color{red}1111$\check1$01}{\color{cyan}\cancel1000}, {\color{red}1111$\check1$01}{\color{cyan}\underline0000}, {\color{red}1111$\check1$01}{\color{cyan}$\check0$1000}.
We can simply let $p_i=k+10$.
		
		\item $\mathbf y[k+4:k+11]=11111011$. 
It results from only one error pattern of
		{\color{red}$\mathbf a^{(i)}$}:	 {\color{red}1111$\check1$01}, leading to $p_i=k+10$.
		
		\item $\mathbf y[k+4:k+11]=11111100$. 
It results from three error patterns of
		{\color{red}$\mathbf a^{(i)}$}{\color{cyan}$\mathbf b^{(i+1)}v_1^{(i+1)}$}: {\color{red}1111\cancel01}{\color{cyan}1000}, {\color{red}1111\underline01}{\color{cyan}\cancel1000}, {\color{red}1111\underline11}{\color{cyan}\underline0000}.
We can simply let $p_i=k+8$. 
		
		\item $\mathbf y[k+4:k+11]=11111101$.
It results from three error patterns of 
		{\color{red}$\mathbf a^{(i)}$}{\color{cyan}$\mathbf b^{(i+1)}v_1^{(i+1)}$}: {\color{red}1111\cancel01}{\color{cyan}10$\check1$00}, {\color{red}1111\cancel01}{\color{cyan}10\underline10}, {\color{red}1111\underline11}{\color{cyan}$\check0$1000}.
We can make use of $\mathbf y[2k+13:2k+18]$ to determine $p_i$ similarly as in Case B3).
		
		\item $\mathbf y[k+4:k+11]=11111110$.
It results from three error patterns of
		{\color{red}$\mathbf a^{(i)}$}{\color{cyan}$\mathbf b^{(i+1)}v_1^{(i+1)}$}: {\color{red}1111\cancel01}{\color{cyan}1\underline100}, {\color{red}1111\cancel01}{\color{cyan}1$\check1$000}, {\color{red}1111\underline11}{\color{cyan}1000}.
We can make use of $\mathbf y[2k+13:2k+18]$ to determine $p_i$ similarly as in Case B3).
		
		\item $\mathbf y[k+4:k+11]=11111111$.
It results from three error patterns of
		{\color{red}$\mathbf a^{(i)}$}{\color{cyan}$\mathbf b^{(i+1)}v_1^{(i+1)}$}: {\color{red}1111\underline11}{\color{cyan}1\underline100}, {\color{red}1111\underline11}{\color{cyan}1$\check1$000}.
		We can simply let $p_i=k+9$.
	\end{enumerate}
	In summary, for each case in Table \ref{AllCases-edit}, we can obtain $p_i$ such that
	$\mathbf y[p_i+1:\infty]\in\mathcal B_{\textrm{seg}}^{\textrm{single-edit}}(\mathbf s^{(i+1)}\mathbf s^{(i+2)}\cdots\mathbf s^{(t)})$.
This completes the proof.

}

%\bibliographystyle{IEEEtran}
%\bibliography{reference}

\end{document}